\documentclass[11pt,a4paper]{article}
\usepackage{jheppub}

\usepackage{graphicx}
\usepackage{amsmath,amsthm,amssymb,bm,amsfonts}

\usepackage{color}
\usepackage[makeroom]{cancel}
\usepackage[normalem]{ulem}

\def\kb{\bm{k}}

\definecolor{greeen}{rgb}{0.1,0.5,0.2}
\definecolor{grey}{rgb}{0.7,0.7,0.7}

\newcommand{\bea}{\begin{eqnarray}}
\newcommand{\eea}{\end{eqnarray}}
\newcommand{\be}{\begin{equation}}
\newcommand{\ee}{\end{equation}}

\title{
BFKL Pomeron loop contribution in diffractive photoproduction and inclusive hadroproduction of $J/\psi$ and $\Upsilon$ }
\author[a]{Piotr Kotko}
\author[b]{Leszek Motyka}
\author[b]{Mariusz Sadzikowski}
\author[c]{Anna M. Sta\'sto}

\affiliation[a]{Institute of Nuclear Physics Polish Academy of Sciences\\ E. Radzikowskiego 152, 31-342 Krak\'ow, Poland}
\affiliation[b]{Institute of Physics, Jagellonian University\\ S. \L{}ojasiewicza 11, 30-348 Krak\'ow, Poland}
\affiliation[c]{Department of Physics, The Pennsylvania State University\\ University Park, PA 16802, U.S.A.}

\abstract{
We analyze contributions to the heavy vector meson production with large transverse momentum in proton--proton and diffractive photon--proton scattering driven by an exchange of two Balitsky--Fadin--Kuraev--Lipatov Pomerons in the squared amplitudes. The Pomerons couple to a single parton and form a Pomeron loop closed by the vector meson impact factors. For the photon--proton case the diffractive cut of the Pomeron loop contributes, and for the inclusive hadroproduction one finds the loop with two cut Pomerons. We  compute both of these Pomeron loop contributions and study in detail their properties. The results are  then used to  calculate the cross sections for diffractive $J/\psi$ photoproduction with large transverse momentum at HERA and the correlated two Pomeron contribution for inclusive $J/\psi$ and $\Upsilon$ production cross sections at the LHC. Within a unified approach a good description of the photoproduction data is found, but correlated two Pomeron mechanism gives only a small contribution to hadroproduction of the vector mesons at the LHC. }


\begin{document}
\maketitle

\section{Introduction}

The heavy vector mesons with negative $C$-parity --- charmonia and bottomonia --- are classical probes of the QCD exchange at high energies. The signals of $J/\psi$, $\psi$ and $\Upsilon$ mesons  in their leptonic decay channels are clear and allow for accurate measurements of the corresponding differential cross sections. The underlying production dynamics is driven by gluonic degrees of freedom and their QCD evolution. The currently accepted pictures of the heavy vector meson production mechanisms in diffractive photon--hadron and 
inclusive
hadron--hadron collisions are, however, quite different. The diffractive photoproduction data at high energies and large transverse momentum obtained by H1 \cite{Aktas:2003zi} and ZEUS  \cite{Chekanov:2002rm} collaborations at HERA have been successfully described \cite{Bartels:1996fs,Forshaw:2001pf,Enberg:2002zy,Enberg:2003jw,Poludniowski:2003yk} assuming an exchange of the non-forward Balitsky--Fadin--Kuraev--Lipatov (BFKL) Pomeron \cite{Fadin:1975cb,Kuraev:1977fs,Balitsky:1978ic,Bronzan:1977yx,Bronzan:1977ag,Jaroszewicz:1980mq,Lipatov:1985uk,Fadin:2004zq,Fadin:2005zj,Lipatov:1996ts} in the diffractive amplitude. On the other hand, in the high energy inclusive hadroproduction of $J/\psi$ and $\Upsilon$, a good description of data from hadron colliders  requires adopting the Color Octet Model (COM) \cite{Bodwin:1994jh,Cho:1995vh,Cho:1995ce,Petrelli:1997ge,Nayak:2005rt,Butenschoen:2009zy,Butenschoen:2010rq,Butenschoen:2011yh}, see also Refs.\ \cite{Lansberg:2019adr,Lansberg:2006dh} for a review. In the present paper we shall investigate the diffractive photoproduction and a contribution to the inclusive hadroproduction of heavy vector mesons assuming the same underlying QCD dynamics of two BFKL Pomeron exchange.

The standard picture of diffractive photoproduction of vector mesons at HERA at large momentum transfer assumes exchange of gluonic hard color singlet across large rapidity distance between the incoming photon --- outgoing meson vertex and the  diffractive remnant of the proton. The kinematics of this process allows to apply the high energy limit in which the dominant contribution to the color singlet amplitude is given by the BFKL Pomeron \cite{Balitsky:1978ic,Kuraev:1977fs,Lipatov:1985uk,Bronzan:1977yx,Bronzan:1977ag,Jaroszewicz:1980mq}. By the BFKL Pomeron one understands the system of two Reggeized gluons  in the $t$-channel interacting by exchange of usual gluons. The propagation of the Reggeized gluons and the effective interactions between them are derived in QCD in 
the
high energy limit. In more detail, the exchange amplitude is described by the BFKL evolution equation that formally resums logarithmically enhanced perturbative corrections to all orders. In the BFKL approach one resums logarithms of a ratio of a large collision energy $\sqrt{s}$ and other, much smaller mass scales e.g.\ the meson mass or the momentum transfer. These logarithms are  related to the rapidity distance $Y$ between the projectile and the target in the high energy scattering process. So far the BFKL resummation in QCD  was performed at the leading logarithmic (LL) \cite{Balitsky:1978ic,Kuraev:1977fs,Lipatov:1985uk,Bronzan:1977yx,Bronzan:1977ag,Jaroszewicz:1980mq} and next-to-leading logarithmic (NLL) approximation \cite{Fadin:1997hr,Fadin:1998py,Camici:1996st,Camici:1997ij,Ciafaloni:1998gs,Fadin:2004zq,Fadin:2005zj}.  In the LL approximation one resums 
terms $\delta ^{(n)} _{\mathrm{LL}} \sim \alpha_s^n Y ^n$ 
while
$\delta ^{(n)} _{\mathrm{NLL}} \sim \alpha_s^{n+1} Y ^n$ terms are resummed by the NLL BFKL evolution. 
The BFKL formalism assumes high energy (or $k_T$) factorization in which hard matrix elements are factorized in rapidity space from the BFKL  evolution. In addition, matrix elements are off-shell, with initial quarks or gluons carrying non-zero transverse momenta unlike in the standard collinear approximation.  Hence also the BFKL Pomeron may carry non-zero transverse momentum, and the corresponding amplitude is governed by the non-forward BFKL Pomeron. This formalism  was applied \cite{Ginzburg:1996vq,Forshaw:1995ax,Motyka:2001zh,Enberg:2002zy,Enberg:2003jw} some time ago to the data from HERA on $J/\psi$, $\rho$ and $\phi$ diffractive mesons photoproduction \cite{Chekanov:2002rm,Aktas:2003zi,Chekanov:2009ab} with large transverse momentum $p_T$ and was shown to describe the data well. 
In this paper we revisit the diffractive $J/\psi$ photoproduction at HERA and use the established description of this process to estimate the BFKL Pomeron loop contribution to inclusive vector meson hadroproduction.

	The  COM
	 of inclusive heavy meson hadroproduction assumes non-zero amplitudes for a change of the quantum numbers (in particular color and angular momentum) between the partonic phase and the meson \cite{Bodwin:1994jh,Cho:1995vh,Cho:1995ce}. More specifically, in the partonic subprocess heavy quark--antiquark pair $Q\bar Q$ is produced with an arbitrary color and angular momentum quantum numbers, and the transition to the final state meson is governed by separate multiplicative coefficients for each set of the partonic quantum numbers. The values of these coefficients are obtained by fitting the predicted cross sections to experimental data. The theoretical basis for this mechanism comes from two complementary sources. First, within heavy quark effective theory one finds non-zero amplitudes of higher Fock states in the heavy vector meson wave function \cite{Bodwin:1994jh}, for instance a state $Q\bar Q g$ with an additional gluon $g$. Obviously, the quantum numbers of $Q\bar Q$ in this state do not match the quantum numbers of the meson. From another perspective, before the partonic $Q\bar Q$ makes the meson a process of hadronization occurs, in which the quantum numbers may change. This is a picture of fragmentation of the primordial $Q\bar Q$ state into the heavy meson \cite{Nayak:2005rt}. The hadronization process does not have satisfactory perturbative description as it occurs at low (hadronic) scales and depends on non-perturbative properties of the QCD vacuum. In both 
	  scenarios the transition amplitudes of $Q\bar Q$ to the meson cannot be derived from theory  --- only order-of-magnitude estimates can be obtained. Nevertheless the COM has a solid theoretical basis, and it is strongly supported by the successful fits of its predictions to the bulk of experimental data. However, since the parameters of the model are fitted, there is still room for other than COM possible mechanisms of the inclusive vector meson hadroproduction.

	The classical alternative to the  COM
	 of the heavy meson production is the Color Singlet Mechanism (CSM) \cite{Baier:1982zz,Baier:1983va,Glover:1987az}. In fact, the CSM was considered to be the standard QCD prediction before it was contradicted by the Tevatron data  \cite{Abe:1992ww,Abe:1997yz}. In this approach one assumes the exact matching of the quantum numbers of the produced  partonic $Q\bar Q$ state and the final state meson. The main advantage of this mechanism is its completeness within perturbative QCD and no need for additional parameteres. For the $C$-odd mesons $V$ at the leading twist the CSM is driven by the $g + g \to V+g$ partonic scattering. The predictions of the CSM, however, fail badly in describing the $J/\psi$ hadroproduction at the Tevatron and the LHC, see e.g.\ \cite{Lansberg:2019adr,Lansberg:2006dh}. For the total inclusive vector meson hadroproduction cross sections the CSM both at the leading order (LO) and at the next-to-leading order (NLO) are more than one order of magnitude below the data \cite{Lansberg:2019adr,Lansberg:2006dh}. Moreover, the CSM predictions lead to  the distribution in the meson transverse momentum $p_T$ which is much too soft, while the COM is able to describe well the $p_T$ dependence of the meson production cross section. The CSM and COM approaches were also extended from the collinear approximation to the $k_T$-factorization framework \cite{Hagler:2000dd,Hagler:2000eu,Yuan:2000cp,Baranov:2002cf,Saleev:2003ys,Kniehl:2006sk,Cisek:2017gno}. It was found that also in the $k_T$-factorization approach the CSM model is much below the data for the direct $J/\psi$ production at large transverse momenta.

	Beyond the leading twist approximation the  CSM
	 may be realized also with a fusion of three initial state gluons. At the partonic level the meson formation occurs by $g+g+g \to V$ diagrams with the coupling through the heavy quark loop. The Three Gluon Fusion (3GF) mechanism was considered in Ref.~\cite{Alonso:1989pz} as a contribution to $J/\psi$ hadroproduction, and in Ref.\ \cite{Khoze:2004eu} it was proposed as a possible leading contribution to heavy vector meson hadroproduction. Then it was further studied in Refs.\ \cite{Motyka:2015kta,Kopeliovich:2017jpy,Levin:2018qxa}. 
Moreover, recently contributions of multiple gluon couplings in the $J/\psi$ production were combined with the COM \cite{Ma:2015sia,Ma:2018qvc}. In this approach a good description of the $J/\psi$ hadroproduction data was obtained, including the meson polarization. Although the uncorrelated 3GF mechanism  is an implicit contribution of this framework, at larger $p_T$ the cross section is dominated by the COM contributions.

In the 3GF mechanism one of the gluons comes from one hadronic beam (the projectile), and two other ones from the other beam (the target). These two gluons in the $t$-channel can be taken either as completely independent (uncorrelated) or as coming from a single parton of the target (correlated). The two scenarios correspond to an uncorrelated double gluon distribution in the target, and to the correlated contribution in the double gluon distribution, respectively. A detailed study of the uncorrelated contribution to the 3GF mechanism showed that it may contribute to the total $J/\psi$ hadroproduction cross section as a fraction of about 20 -- 25\% \cite{Motyka:2015kta}. The obtained $p_T$-dependence of the meson production differential cross section $d\sigma / dp_T$ was found to be much steeper than the experimental data. Specifically, at larger values of $p_T$ the experimental data can be approximated with a power law:  $d\sigma / dp_T \sim 1/p_T ^n$  with $n \simeq 5$, while with the uncorrelated 3GF mechanism $n>8$ is obtained. This rather steep $p_T$ dependence is well understood as the uncorrelated 3GF contribution enters at a higher twist, and it is suppressed at large $p_T$ by an additional factor of $(\Lambda_h / p_T)^2$ with respect to the leading twist, where $\Lambda_h$ is a small hadronic scale.

	In this paper we analyze in detail the correlated 3GF mechanism. The two gluons that enter the vector meson production vertex from the target side are assumed to come from a splitting of a single parton: quark or gluon in the target. Since the parent parton is point-like, it does not introduce any additional hadronic scale and one expects a harder dependence of the meson $p_T$-distribution than it was for the uncorrelated 3GF. At the lowest order this mechanism occurs through  $g + g \to V + g$ (or $g + q \to V + q$) partonic process with an exchange of two gluons between the $g \to V$ transition vertex at the side of the projectile and the $g \to g$ (or $q \to q$) scattering at the target side.  At the amplitude level the two gluon exchange in the $t$-channel carries the symmetric color octet. The important feature of the process at the lowest order is a flat dependence on the rapidity distance $Y$ between the projectile gluon and the target gluon. Beyond the lowest order approximation the amplitude of this subprocess is modified by QCD radiative corrections. These corrections can be resummed 
by a QCD evolution equation. In the high energy limit corresponding to $Y\gg 1$, the cross section of correlated 3GF cross section is driven by an exchange of four Reggeized gluons in the total color singlet state that interact with BFKL kernels. This system is described by Bartels--Kwieci\'{n}ski--Prasza\l{}owicz (BKP) evolution equation \cite{Bartels:1978fc,Bartels:1980pe,Kwiecinski:1980wb}.
	It was shown in Ref.\ \cite{Bartels:1993it} that in the large $N_c$ limit the system of four Reggeized gluons in the color singlet channel may be approximated by two independent BFKL Pomerons. In consequence one expects a strong enhancement $\sim \exp(2\Delta_{\mathbb{P}}) $ of the cross section at large $Y$ by the double BFKL evolution with the BFKL Pomeron intercept $\Delta_{\mathbb{P}}  \sim 0.3$. In addition, the anomalous dimensions of the BFKL Green's functions could lead to a less steep $p_T$ dependence. 
	So, despite this contribution enters at the ${\cal O} (\alpha_s^5)$ order and is a well defined part of the NNLO correction to the CSM contribution, it may be important due to strong effects of the BKP evolution. The first  estimate of this contribution \cite{Khoze:2004eu} suggested that it may reproduce the inclusive $J/\psi$ hadroproduction cross section data from the Tevatron. In this paper we perform a detailed calculation of the correlated 3GF cross section to verify its importance.

	In the analysis we shall use a connection between the correlated 3GF mechanism and the diffractive photoproduction of vector mesons at large $p_T$. This connection originates from the kinematical identity of the impact factors corresponding to $\gamma + (2g) \to V$ and $g + (2g) \to V$ transitions. At the leading order the difference between these two impact factors comes only from the color factors and the gauge coupling constants. Hence, the cross sections for 
	a vector meson production by gluon scattering
	 off a partonic target and for the meson diffractive photoproduction are proportional to each other. This relation imposes important constraints on the correlated 3GF mechanism. When the BKP/BFKL evolution is taken into account however, the diffractive photoproduction and the 3GF mechanism are different due to the different color flow. The diffractive photoproduction cross section corresponds to the diffractive cut of two BFKL Pomeron exchange, and the correlated 3GF cross section to an exchange of two cut BFKL Pomerons. Thus independent evaluation is necessary for these two processes. In both 
	  cases, however, at the level of cross section one finds the topology of the BFKL Pomeron loop spanned between the meson impact factors at the projectile side and the parton impact factor at the target side. So, besides evaluating the correlated 3GF contribution with the BKP/BFKL evolution effects included, we shall also revisit the diffractive photoproduction case and analyse in detail the properties of the BFKL Pomeron loop in the $t$-channel. \\

	The paper is organized as follows. In the next section  we describe the theoretical framework for the diffractive heavy vector meson production in DIS,
	and  the non-forward BFKL evolution. In Section~\ref{sec:Hadroproduction} the two-Pomeron contribution to the vector meson hadroproduction is analyzed. The details of the color factors are discussed and the BFKL Pomeron  in the conformal representation is described.
	In Section~\ref{sec:PomLoopParton} the properties of the Pomeron loop
	at the parton level are analyzed, in particular the dependence on the transverse momentum. In Section~\ref{sec:Results}
	we present the comparison of our numerical calculations with the vector meson production in diffractive DIS as well as in hadroproduction, and we discuss the results in Section\ \ref{sec:Discussion}. Finally, in Section \ref{sec:Summary} we give summary and conclusions.

\section{Diffractive heavy vector meson production}
\label{sec:Diff}

We begin with a short recollection of the perturbative QCD approach to the hard color singlet exchange in the diffractive heavy vector meson $V$ photoproduction off a proton at large momentum transfer $t$.
The process was investigated in detail at HERA in $e^{\pm}p$ collisions with invariant c.m.s.\ energy  $\sqrt{S}=318 \, {\rm GeV}$. 
In the  measurement of the process $e^{\pm} p \to e^{\pm}  V\, X$, a large rapidity gap devoid of particles is required. It separates the produced vector meson V and the dissociated proton remnant $X$.
The $e^{\pm} p$ cross section may be factorized into a universal flux function of quasi-real photons in the electron and the cross section for the diffractive photoproduction subprocess,

\begin{equation}
\gamma\, p \rightarrow V\, X\, ,
\label{eq:DiffProc}
\end{equation}
with the rapidity gap. In this process the photon--proton invariant mass, $\sqrt{s} = \sqrt{zS}$, is assumed to be much larger than all the other scales present in the process, hence $s\gg |t|$ and $s\gg M_V^2$, where $M_V$ is the meson mass. The applicability of the perturbative QCD is ensured by the conditions $|t|\gg \Lambda^2 _{\mathrm{QCD}}$ and $M_V^2 \gg \Lambda^2 _{\mathrm{QCD}}$. The diffractively produced $C$-odd state $V$ is assumed to be a heavy vector quarkonium. In this work we focus on the $J/\psi$ meson. Since in the available data the photon flux is strongly dominated by very low virtualities $-q^2 = Q^2$ we take the limit $Q^2 \to 0$ in calculations of the QCD amplitudes.

\begin{figure}[h]
\centerline{\includegraphics[width=0.5\columnwidth,angle=0]{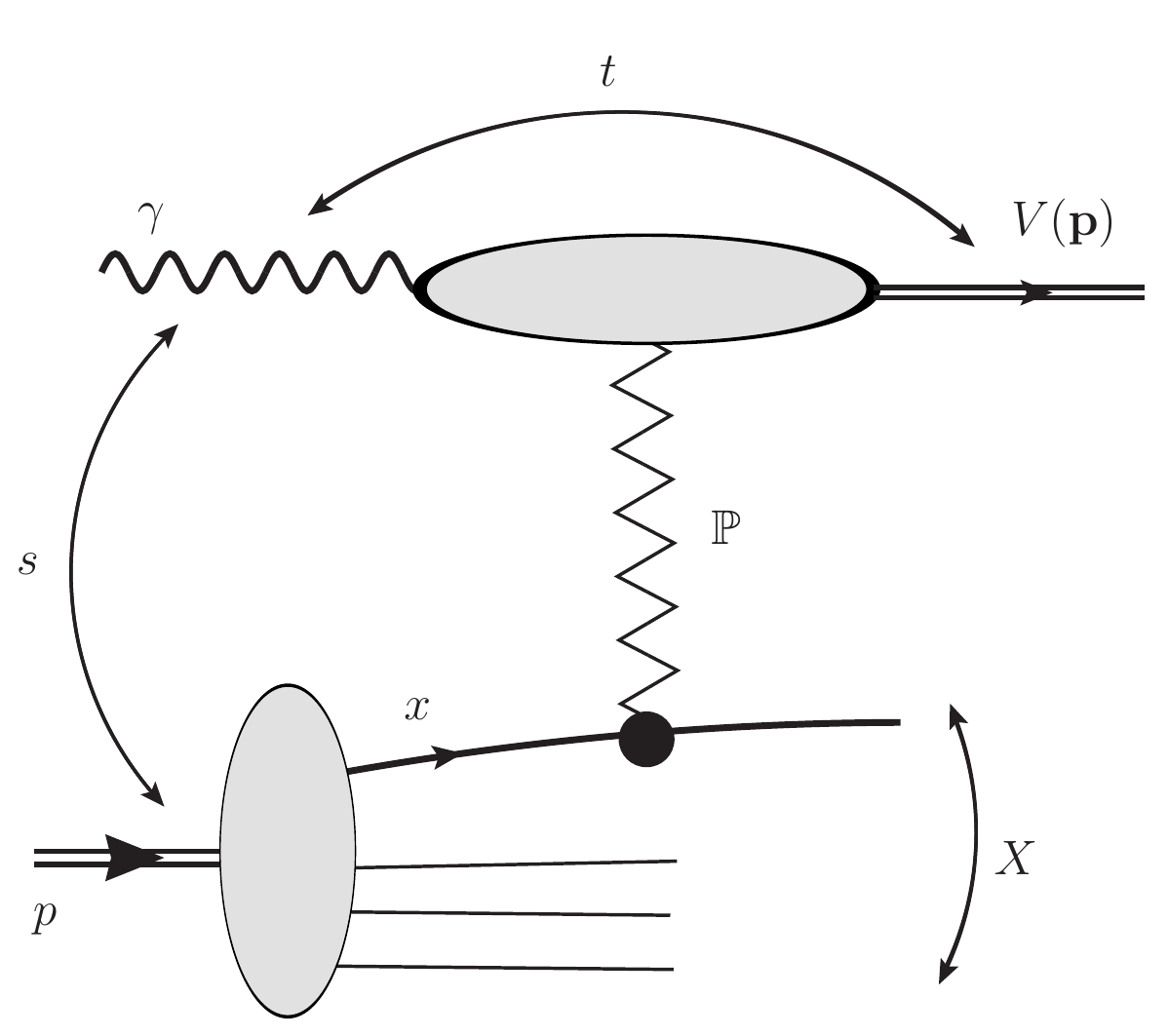}}
\caption{Diffractive photoproduction of a vector meson $V$. The zigzag line corresponds to a perturbative Pomeron coupling to individual partons inside the proton.}
\label{fig:diffraction_1}
\end{figure}

Within the kinematic regime specified above, the color singlet exchange in the $t$-channel of the process \eqref{eq:DiffProc} may be described in QCD as the perturbative Pomeron exchange (Fig.~\ref{fig:diffraction_1}), that is governed by the non-forward BFKL equation \cite{Kuraev:1977fs,Balitsky:1978ic,Bronzan:1977yx,Jaroszewicz:1980mq,Lipatov:1985uk}. 
Due to the large momentum transfer, the cross section can be factorized into a partonic cross section (dominated by the Pomeron exchange) and the collinear parton distribution functions (PDFs), that describe the structure of the proton target:
\begin{equation}
d\sigma = \sum_{i=g,q,\bar{q}} \, \int\! dx \, f_i(x,\mu)\, d\hat\sigma_{\gamma i}(\hat s,t,\mu)\; ,
\label{eq:DiffXsec1}
\end{equation}
where $\hat s =xs$ is the photon--parton invariant mass squared, and $f_i$ is the PDF for the parton $i$ which may be a quark (anti-quark) or a gluon. A natural choice for the factorization  scale is $\mu\sim \sqrt{|t|}$. Here $x$ is the longitudinal fraction of the proton light cone momentum carried by the quark or gluon. In what follows, 
 the transverse two-vectors in the light cone basis are denoted by the bold characters; for example, we denote the vector meson transverse momentum as $\bm{p}$. In the high energy kinematics we have $t\simeq -|\bm{p}|^2 \equiv -p^2$. 
In Eq.\ \eqref{eq:DiffXsec1} the coupling of the high-$t$ Pomeron to the proton is assumed to occur through coupling to the individual partons. This approximation  was studied in detail and motivated in \cite{Mueller:1992pe,Bartels:1995rn,Motyka:2001zh}.

\begin{figure}[h]
\centerline{\includegraphics[width=7cm,angle=0]{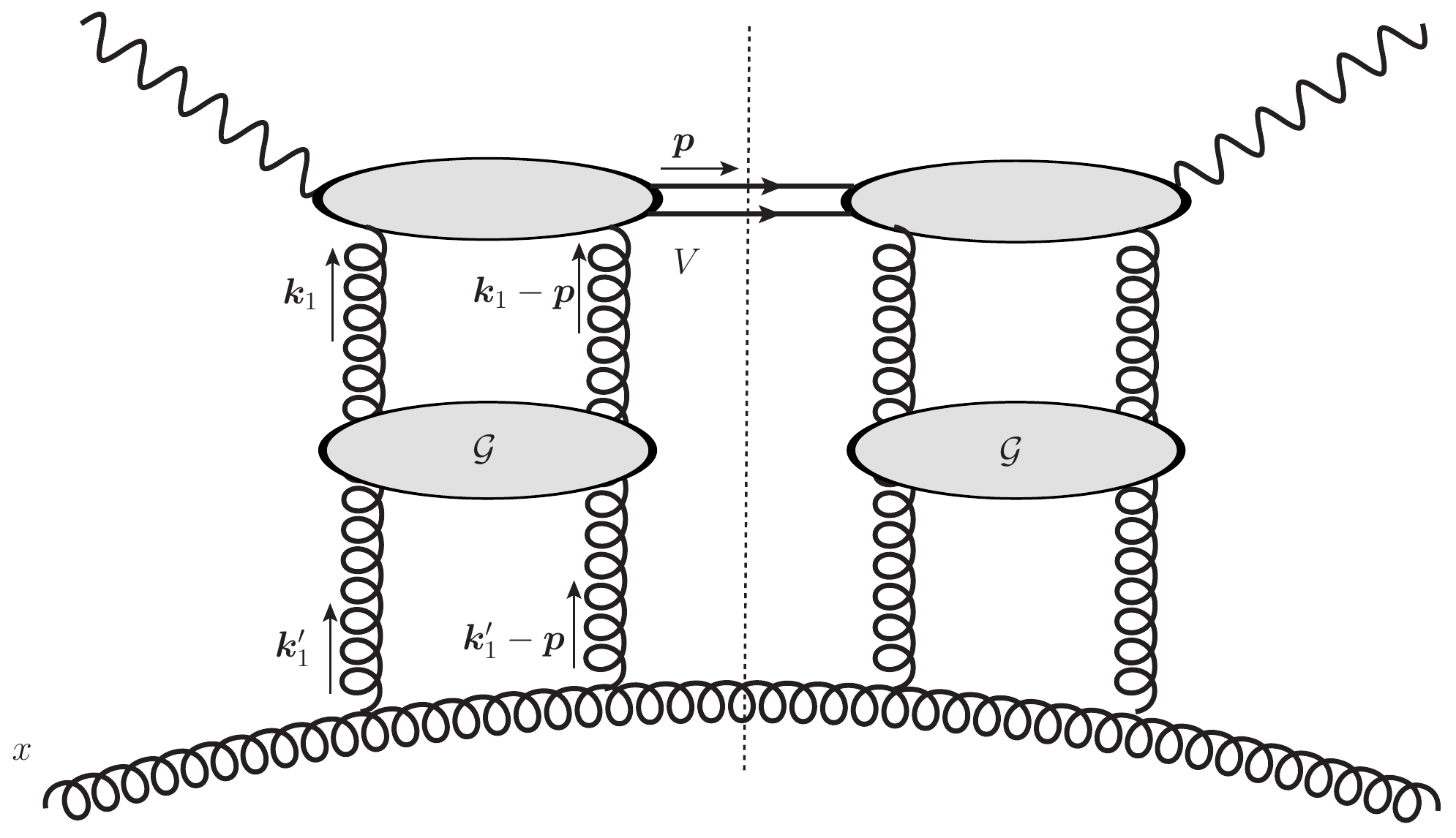}\hspace*{1cm}\includegraphics[width=7cm,angle=0]{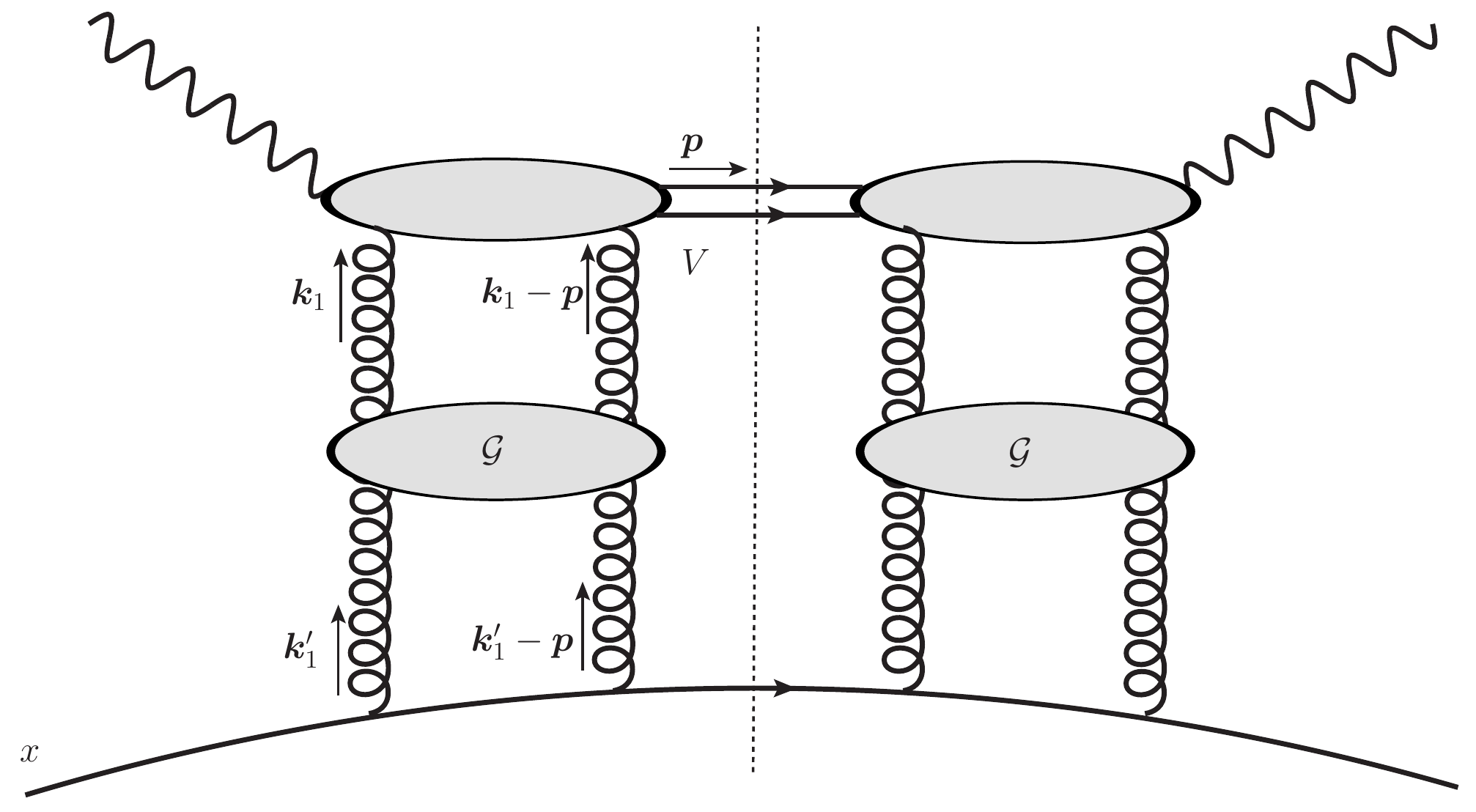}}
\caption{The diagrams contributing to the cross section for diffractive vector meson $V$ photoproduction off a parton: a gluon (left) and a quark (right). The non-forward BFKL Green's function $\mathcal{G}$ is evolved along the rapidity gap between the vector meson and the proton remnants. }
\label{fig:diffraction_2}
\end{figure}

In the high energy limit, the kinematic part of the BFKL Pomeron coupling to quarks and gluons is the same.
 The only difference between the quark and gluon partonic target comes from the color factors,
\begin{equation}
d\hat \sigma_{\gamma i} = C_{\gamma i} \,\, d\sigma_{\mathtt{1-}\mathbb{P}}\; ,
\label{eq:PartonicXsec1}
\end{equation}
where $C_{\gamma i}$, $i=q,g$, is the color factor and 
\begin{equation}
d\sigma_{\mathtt{1-}\mathbb{P}} = \frac{1}{16\pi\hat s^2}\,\, \left| \mathcal{A}\left(\hat s,t=-p^2 \right) \right|^2\, \frac{d^2\bm{p}}{\pi}\, ,
\label{eq:PartonicXsec2}
\end{equation}
with $\mathcal{A}$ being the amplitude to produce the vector meson through a single Pomeron exchange (Fig.~\ref{fig:diffraction_2}). 
The amplitude is dominated by the imaginary part, and the correction coming from the real part enters only at a subleading order in the logarithmic expansion and may be neglected.  The imaginary part of the amplitude reads 

\begin{equation}
\mathrm{Im}\,  \mathcal{A}\left(\hat s,t=- p^2 \right)
 = \hat s \int\! \frac{d^2\bm{k}_1}{2\pi} \,
\frac{\Phi_{V}\left(\bm{k}_1,\bm{p}\right)
\Phi_{q}\left(y,\bm{k}_1,\bm{p}\right)}
 {(\bm{k}_1 ^2+s_0) \left[(\bm{p}-\bm{k}_1)^2+s_0\right]}\, ,
\label{eq:DiffAmp} 
\end{equation}
where $\bm{k}_1$ and $\bm{p}-\bm{k}_1 \equiv \bm{k}_2$  are transverse momenta of the exchanged gluons, and $\bm{p}$ is the transverse momentum carried by the Pomeron.
$\Phi_{V}$, $\Phi_{q}$ are the  impact factors for the vector meson and for the quark, respectively. They are stripped off
 the color factors. In addition, the quark impact factor $\Phi_q$ is evolved to rapidity $y$ using the BFKL evolution.
 The exact form of the impact factors will be described below in this section. 
  The rapidity evolution length $y$ of the quark impact factor is defined by the relation
\begin{equation}
y = \log\left(\frac{\hat s}{\Lambda^2}\right) \, ,
\label{eq:rap1}
\end{equation}
where we set $\Lambda = E_T \equiv \sqrt{M_V^2 + \bm{p}^2}$. This choice is different than the choice made in earlier studies \cite{Forshaw:2001pf,Enberg:2002zy,Enberg:2003jw,Poludniowski:2003yk}, where $\Lambda = M_V$ was used. The latter value was selected as a result of fits driven mostly by the light vector meson high-$t$ photoproduction data. The $J/\psi$ data, however, were well described both with $\Lambda = E_T$ and  $\Lambda = M_V$. Hence in this paper we use $\Lambda = E_T$, the value with a straightforward kinematic motivation. This choice was used e.g.\ in Ref.\ \cite{Bartels:1996fs}.

 The small parameter $s_0$ is a phenomenological infrared cutoff that mimicks the effects of 
 the  color confinement; the results are, however, finite for $s_0\rightarrow 0$. We introduce this parameter following the approach proposed in Ref.\ \cite{Kwiecinski:1998sa}. 
The  amplitude does not depend on the produced meson direction in the transverse plane, hence the phase space $d^2\bm{p}$ in (\ref{eq:PartonicXsec2}) may be trivially integrated over the azimuthal angle. We present the results in this form to keep the notation uniform with the more complicated case of the vector meson hadroproduction, which we shall describe in Section~\ref{sec:Hadroproduction}.

In the calculations we take the non-relativistic approximation for the meson wave function. With this assumption the lowest order photon to vector meson impact factor reads \cite{Ryskin:1992ui,Ginzburg:1996vq,Bzdak:2007cz}
\begin{equation}
\Phi^{ab}_{\gamma V} \left(\bm{k}_1,\bm{p}\right)  =   \Phi_{V}\left(\bm{k}_1,\bm{p}\right)\, {\delta^{ab} \over N_c}\,,
\label{eq:Phi_gamV}
\end{equation}
where $a$, $b$ are color indices of the exchanged gluons, and the kinematic part of the impact factor reads
\begin{equation}
\Phi_{V}\left(\bm{k}_1,\bm{p}\right) = 
16\pi e e_q \alpha_s M_V g_{V}  \left[\frac{1}{M_V^2+\bm{p}^2}-\frac{1}{M_V^2+(\bm{p}-2\bm{k}_1)^2}\right] \, ,
\label{eq:Phi_V}
\end{equation}
where 
\begin{equation}
g_{V} = \sqrt{\frac{3 M_V\Gamma_{V\rightarrow ll}}{16\pi\alpha^2 _{em} e_q^2}}\; ,
\label{eq:gJpsi}
\end{equation}
with $e_q$ being the charge of the quark in the meson in units of the elementary charge $e$, $M_V$ -- the mass of the vector meson, and $\Gamma_{V\rightarrow ll}$ its leptonic decay width. 
The photon to vector meson impact factor \eqref{eq:Phi_V} is valid for the transverse polarizations of the photon and of the vector meson. The incoming photon is quasi-real hence its polarization is constrained to be transverse. The amplitude of the transverse photon to longitudinaly polarized vector meson was estimated in Ref.\ \cite{Ginzburg:1996vq} to be small. Therefore the amplitude described by \eqref{eq:Phi_V} provides the dominant contribution
to diffractive vector meson photoproduction.

The quark impact factor in the color singlet channel $\Phi^{\bm{1},ab}_{q}\left(y,\bm{k}_1,\bm{p}\right)$ is also factorized into the color part and the kinematic part,
\begin{equation}
\Phi^{\bm{1},ab}_{q}\left(y,\bm{k}_1,\bm{p}\right) = \Phi_{q}\left(y,\bm{k}_1,\bm{p}\right) \, {\delta^{ab} \over 2 N_c}\,, 
\label{eq:Phi_qq}
\end{equation} 
where the notation is the same as in the case of the photon--vector meson impact factor.
The diffractive gluon impact factor $\Phi^{\bm{1},ab}_{g}$ differs from the diffractive quark impact factor only by the color factor,
\begin{equation}
\Phi^{\bm{1},ab}_{g}\left(y,\bm{k}_1,\bm{p}\right) = \Phi_{q}\left(y,\bm{k}_1,\bm{p}\right) \, {N_c \over N_c^2 -1}\, \delta^{ab}\, .
\label{eq:Phi_gg}
\end{equation}

The kinematic part of the quark impact factor $\Phi_q\left(y,\bm{k}_1,\bm{p}\right)$ in \eqref{eq:DiffAmp} is the solution of the non-forward BFKL equation with the initial condition given by
the leading order quark impact factor  \cite{Kwiecinski:1998sa}
\begin{equation}
\Phi_{q,0}(\bm{k}_1, \bm{p}) = \alpha_s \, .
\label{eq:Phiq0}
\end{equation}
At a given momentum transfer $\bm{p}$ through the Pomeron, the evolved quark impact factor $\Phi_{q}\left(y,\bm{k}_1,\bm{p}\right)$ may be represented as convolution of the leading order impact factor $\Phi_{q,0}$ with the non-forward BFKL Green's function $\mathcal{G}_y$:
\begin{equation}
\Phi_q\left(y,\bm{k}_1,\bm{p}\right) = 
 \int\! d^2\bm{k}'_1 \, \Phi_{q,0}(\bm{k}'_1, \bm{p})\, 
  \mathcal{G}_y(\bm{k}_1,\bm{k}' _1;\bm{p}) \, .
 \label{eq:Phiq1}
\end{equation}

In what follows, we shall use transverse momentum variables $\bm{k}$, $\bm{k}'$ that reflect the symmetry of the problem: $\bm{k} = (\bm{k}_2 - \bm{k}_1)/2$, $\bm{k}' = (\bm{k}'_2 - \bm{k}'_1)/2$. The transverse momenta of gluons
 take the form 
\begin{equation}
\bm{k}_{1} = \frac{\bm{p}}{2} - \bm{k},\quad
\bm{k}_{2} = \frac{\bm{p}}{2} + \bm{k},\quad
\bm{k}_{1}'= \frac{\bm{p}}{2} - \bm{k}',\quad
\bm{k}_{2}'= \frac{\bm{p}}{2} + \bm{k}'.
\end{equation}
The explicit form of the leading logarithmic BFKL equation that defines the quark impact factor $\Phi_q\left(y,\bm{k}_1,\bm{p}\right)$ is the following 
\begin{multline}
\Phi_q\left(y,\bm{k}_1,\bm{p}\right) = \Phi_{q,0}\left(\bm{k}_1,\bm{p}\right)
 + \bar{\alpha}_s \,  \int_0^y\!dy'\int\!\frac{d^2\bm{k}'}{2\pi}
\frac{1}{\left(\bm{k}'-\bm{k}\right)^2+s_0} \\
\Bigg\{\left[ \frac{\bm{k}_1^2}{\bm{k}'^{\,2}_1+s_0} + \frac{\bm{k}_2^2}{\bm{k}'^{\,2}_2+s_0}
-p^2 \frac{\left(\bm{k}'-\bm{k}\right)^2+s_0}{\left(\bm{k}'^{\,2}_1+s_0\right)\left(\bm{k}'^{\,2}_2+s_0\right)}\right]
\Phi_q\left(y',\bm{k}' _1,\bm{p}\right) \\
-\left[ \frac{\bm{k}_1^2}{\bm{k}'^{\,2}_1 +\left(\bm{k}'-\bm{k}\right)^2+s_0} 
+ \frac{\bm{k}_2^2}{\bm{k}'^{\,2}_2+\left(\bm{k}'-\bm{k}\right)^2+s_0}\right]
\Phi_q\left(y',\bm{k} _1,\bm{p}\right)\Bigg\}\, ,
\label{eq:BFKL0}
\end{multline}
with $\bar{\alpha}_s=\alpha_s N_c/\pi$.

The non-forward BFKL equation given in \eqref{eq:BFKL0} is solved numerically, using an approximation introduced in \cite{Kwiecinski:1998sa}.
The idea is to use the Fourier decomposition of the impact factor 
 w.r.t.\ the angle $\phi_k$ between $\bm{k}$ and $\bm{p}$,
\begin{equation}
\Phi_q(y,\bm{k}_1,\bm{p}) = \sum_{m=0} ^{\infty} \Phi_q ^{(m)} (y,k^2,p^2) \cos(m \phi_k),
\end{equation}
where the Fourier coefficients
\be
\Phi_q ^{(0)} (y,k^2,p^2) = \int _0 ^{2\pi} {d\phi_ k \over 2\pi}\, \Phi_q (y,\bm{k}_1,\bm{p})\, ,
\ee  and  
\be
\Phi_q ^{(m)} (y,k^2,p^2) = \int _0 ^{2\pi} {d\phi_ k \over \pi}\, \Phi_q (y,\bm{k}_1,\bm{p}) \cos(m\phi_k)\, ,
\ee 
for $m>0$. We have checked that the full solution $\Phi_q (y,\bm{k}_1,\bm{p})$ is well approximated by the leading component $\Phi_q ^{(0)} (y,k^2,p^2)$, in accordance with results \cite{Kwiecinski:1998sa,Enberg:2002zy}.
Since this approximation leads to much greater numerical efficiency, with a negligible effect on accuracy,  we use it in the estimates of the cross section. The leading Fourier component with $m=0$ obeys the equation,
\begin{multline}
\Phi^{(0)} _q\left(y,k^2,p^2\right) = \Phi _{q,0}\left(k^2,p^2\right)
 + \bar{\alpha}_s \, \int\!\frac{d\phi_k}{2\pi} \int_0^y\!dy'\int\!\frac{d^2\bm{k}'}{2\pi}
\frac{1}{\left(\bm{k}'-\bm{k}\right)^2+s_0} \\
\Bigg\{
\left[ \frac{\bm{k}_1^2}{\bm{k}'^{\,2}_1+s_0} + \frac{\bm{k}_2^2}{\bm{k}'^{\,2}_2+s_0}
-p^2 \frac{\left(\bm{k}'-\bm{k}\right)^2+s_0}{\left(\bm{k}'^{\,2}_1+s_0\right)\left(\bm{k}'^{\,2}_2+s_0\right)}\right]
\Phi^{(0)}_q\left(y',k'^{\,2},p^2\right) \\
-\left[ \frac{\bm{k}_1^2}{\bm{k}'^{\,2}_1 +\left(\bm{k}'-\bm{k}\right)^2+s_0} 
+ \frac{\bm{k}_2^2}{\bm{k}'^{\,2}_2+\left(\bm{k}'-\bm{k}\right)^2+s_0}\right]
\Phi^{(0)}_q\left(y',k^2,p^2\right)\Bigg\}\, .
\label{eq:BFKL1}
\end{multline}
The independence of the leading order quark impact factor on the angles allows to set 
$
\Phi _{q,0}\left(k^2,p^2\right) = \Phi _{q,0}\left(\bm{k},\bm{p}\right) = \alpha_s
$. Analogously we define 
$
\Phi _{V} ^{(0)}\left(k^2,p^2\right) = \int_0 ^{2\pi} [d\phi_k/(2\pi)]\, \Phi_V(\bm{k}_1,\bm{p})$. One finds that 
$\Phi _{V} ^{(0)}\left(k^2,p^2\right) = \left.\Phi_V(\bm{k}_1,\bm{p})\right|_{\bm{k_1} = \bm{p}/2 - \bm{k}}.$

Finally, let us note that the cross section (\ref{eq:DiffXsec1}) can be written in terms of (\ref{eq:PartonicXsec2}) as follows:
\begin{equation}
d\hat\sigma = \int\! dx \left\{ C_{\gamma q}\sum_q \left[ f_q\left(x,\mu\right)+ f_{\bar{q}}\left(x,\mu\right)\right]
+ C_{\gamma g}  f_g\left(x,\mu\right)\right\} \, d\sigma_{\mathtt{1-}\mathbb{P}}\left(xs,t \right)\, .
\label{eq:DiffXsec2}
\end{equation} 
Using Eqs.\ (\ref{eq:Phi_gamV}), (\ref{eq:Phi_qq}) and (\ref{eq:Phi_gg}) it is straightforward to obtain the color factors for the diffractive production off the quark and gluon. The results read
\begin{equation}
C_{\gamma q} = \left(\frac{N_c^2-1}{2N_c^2}\right)^2\, , \qquad C_{\gamma g} = 1\,.
\label{eq:DiffColorfact}
\end{equation}
\section{Two Pomeron contribution to heavy vector meson hadroproduction }
\label{sec:Hadroproduction}

\subsection{Direct approach}
\label{sec:HadroDirect}

\begin{figure}[t]
\centerline{\includegraphics[width=7cm,angle=0]{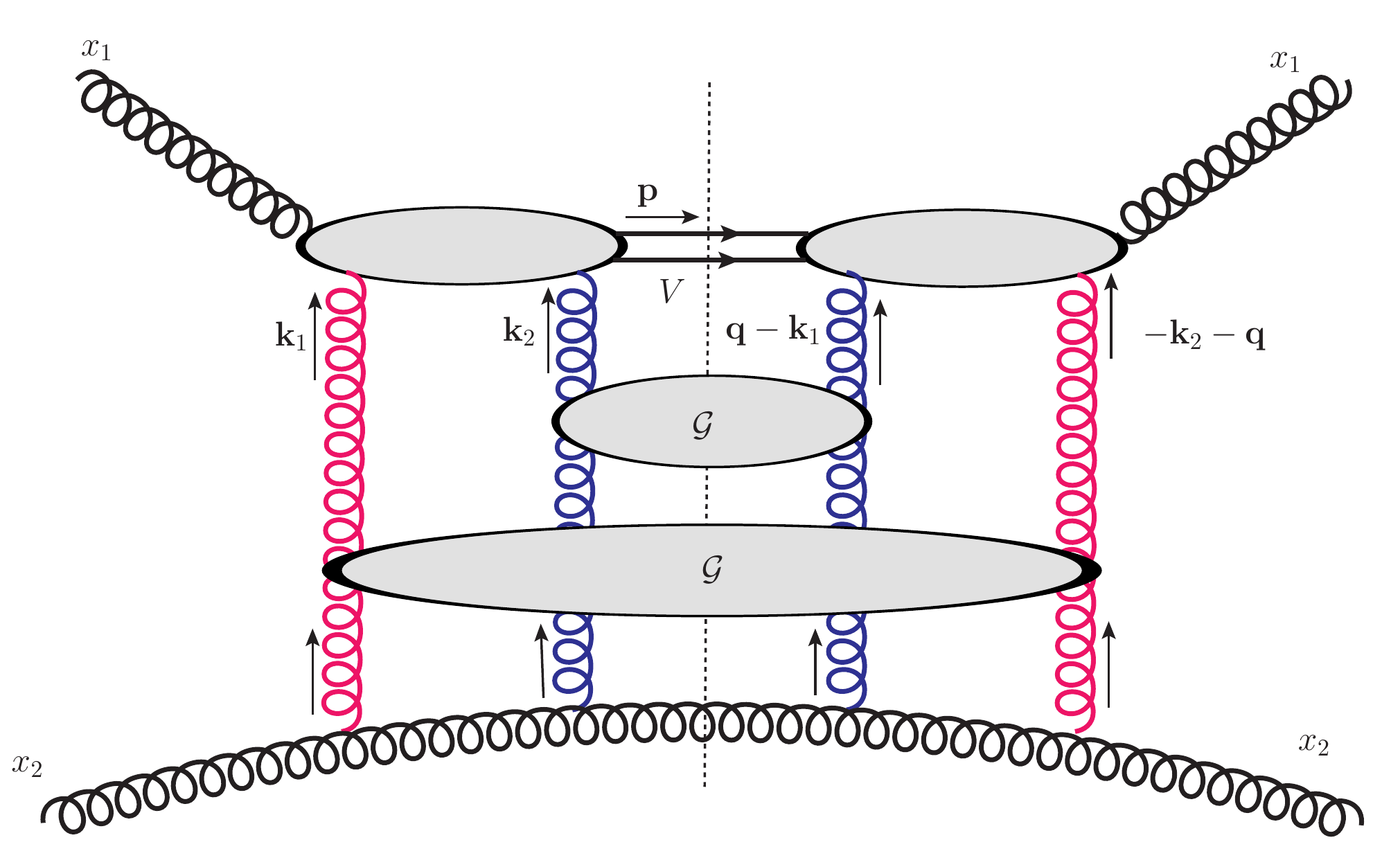}\hspace*{1cm}\includegraphics[width=7cm,angle=0]{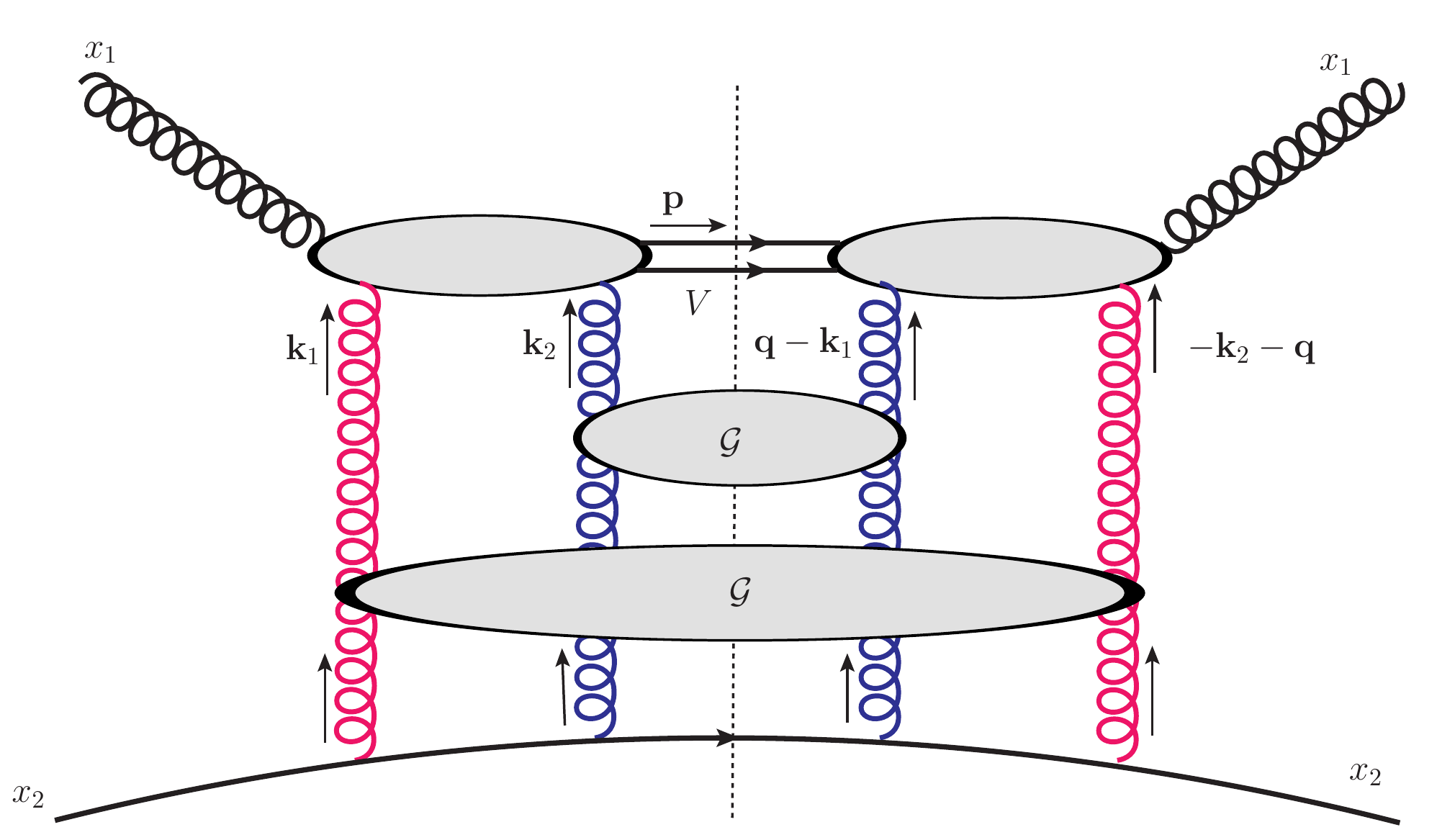}}
\caption{Correlated contributions of $1+2$ gluons fusion to hadroproduction cross section of a heavy vector meson $V$ with partonic targets: a gluon (left) and a quark (right). The blobs with ${\cal G}$ denote the BFKL gluon Green's functions. The two BFKL Green's functions are evolved independently.  Gluons $14$ and $23$ are projected onto color singlet states.}
\label{fig:partoniccrosssection1}
\end{figure}

Let us now consider the heavy vector meson hadroproduction through the (cut) double Pomeron exchange. We focus on a hadronic analogue of the diffractive vector meson photoproduction --- where the projectile is a gluon, and the two Pomerons couple to a single parton in the target. The general diagrams contributing to the partonic cross section, $d\sigma_{\mathtt{2-}\mathbb{P}}$, are depicted in Fig.~\ref{fig:partoniccrosssection1}.
Note that at the lowest orders, the topologies of the considered partonic processes are the same as for the diffractive photoproduction. After inclusion of the evolution, however, the two processes correspond to different cuts through the two Pomerons; 
 one finds the diffractive cut in the photoproduction, and the double cut Pomerons in the hadroproduction. The vector meson impact factor describes the fusion of three gluons into the meson. The coupling of both Pomerons to the single parton in the target leads to a correlation of the gluon distributions in the target, that enter the meson impact factor. This should be contrasted with another possible contribution, where the gluons in the target are uncorrelated, see \cite{Motyka:2015kta}.

In the calculations, the incoming partons from the projectile and target protons  are treated as collinear and
carry longitudinal momentum fractions $x_1$  and $x_2$, whereas the gluon exchange in the $t$-channel is decribed in 
the $k_T$-factorization framework, assuming the high energy limit. 
Hence the vector meson impact factor that enters the calculation describes the transition from the collinear projectile gluon to the vector meson by coupling of two $t$-channel gluons with non-zero transverse momenta. This impact factor may be obtained from the impact factor describing the fusion of three gluons  into the vector meson that was derived in \cite{Bzdak:2007cz}
 by taking the collinear limit 
 for the projectile gluon.  Because of the odd $C$-parity of the meson, the impact factor is fully symmetric in color indices of the gluons, and the functional dependence on the external boson momenta is the same as in the diffractive impact factor for the exclusive vector meson photoproduction. Specifically, in the collinear limit for the projectile gluon, the three-gluon impact factor for inclusive vector meson hadroproduction $\Phi^{a_1 a_2 c} _{gV}$ reads
\begin{equation}
\Phi^{a_1 a_2 c}_{gV} \left(\bm{k}_1,\bm{p}\right)  = {g_s \over e e_q}  \Phi_{V}\left(\bm{k}_1,\bm{p}\right)\, {d^{a_1 a_2 c} \over 2 N_c}\,,
\label{eq:Phi_gV}
\end{equation}
where $g_s$ is the strong coupling constant ($\alpha_s = g_s ^2 / 4\pi$), $d^{a_1 a_2 c}$ is the fully symmetric color tensor, $a_n$ with $n=1,2$, are the color indices of the $t$-channel gluons, and $c$ is the color index of the projectile gluon.

The corresponding leading order impact factors of the target quark ($i=q$) and gluon ($i=g$) may be expressed in the following way:
\begin{equation}
\Phi^{b_1 b_2}_{i,0}\left(\bm{k}_1,\bm{p}\right) = \Phi_{q,0}\left(\bm{k}_1,\bm{p}\right) \, \hat{T}_{R_i} ^{b_2} \hat{T}_{R_i} ^{b_1}\, ,
\end{equation}
where matrices $\hat{T}^{b_n} _{R_i}$ are the generators of the color group in the color charge representation $R_i$ of the target particle. Note that after the projection on the color singlet channel performed by $\mathrm{Tr} (\hat{T}_{R_i} ^{b_2} \hat{T}_{R_i} ^{b_1}) / \mathrm{dim}(R_i)$,  the singlet color impact factors $\Phi_{i,0} ^{\bm{1}, b_1 b_2}$ are recovered.

Using the notation from the previous section we can express the cross section for the inclusive vector meson production in the considered two Pomeron mechanism as 
\begin{multline}
d\sigma = \int\! dx_1 dx_2 \,\, \left\{ \left[ 
f_g\left(x_1,\mu\right) \, d\sigma_{\mathtt{2-}\mathbb{P}}\left(x_1 x_2 S,t,\mu \right) \right. \right. \\
\left. 
\left.
\times \,
\left(
C_{g q}\sum_q 
\left[ f_q\left(x_2,\mu\right)+ f_{\bar{q}}\left(x_2,\mu\right)\right]
+ C_{g g}  f_g\left(x_2,\mu\right)
\right) \right]
 + \left[ x_1\leftrightarrow x_2\right]\, \right\} \, ,
\label{eq:HadroXsec1}
\end{multline} 
where $S$ is the hadronic collision energy squared, the color factors $C_{gq}$ and $C_{gg}$ accommodate the color structure of the vector meson impact factor and the color projection onto the two-Pomeron state. Their calculation is straightforward, but requires certain assumptions on how the projection is made. We shall discuss this issue  later in this section and give the values of the color factors.

The diagrams in Fig.~\ref{fig:partoniccrosssection1}, stripped off the color factors, give the following expression for the partonic cross section for the vector meson production with the transverse momentum~$\bm{p}$: 

\begin{multline}
{d\sigma_{\mathtt{2-}\mathbb{P}}\left(x_1 x_2 S,t,\mu \right) \over d^2\bm{p}}
= {\alpha_s \over 16 \pi^2 \alpha_{em} e_q^2} \int\!\frac{d^2\bm{k}_1}{2\pi} \int\!\frac{d^2\bm{k}_2}{2\pi}
  \int\!d^2\bm{q} \,\, \delta^{2}\left(\bm{k}_1+\bm{k}_2-\bm{p}\right) \\
\times \,
  \frac{
\Phi_V\left(\bm{k}_1,\bm{k}_1+\bm{k}_2\right)\,
\Phi_V\left(\bm{q}-\bm{k}_1,-\bm{k}_1-\bm{k}_2\right)\,
\Phi_q\left(y,\bm{k}_1,\bm{q}\right)\, \Phi_q\left(y,\bm{k}_2,-\bm{q}\right)}
  {\left[\left(\bm{k}_1-\bm{q}\right)^2+s_0\right]
  \left[\left(\bm{k}_2+\bm{q}\right)^2+s_0\right]
\left(k_1^2+s_0\right)\left(k_2^2+s_0\right)
} 
\, ,
  \label{eq:HadronPartonicXsec}
\end{multline}
where we set the rapidity evolution length to $y = \log(x_1 x_2 S / (M_V ^2 + p^2))$, and the remaining notation is the same as in  the previous section.
In fact, the above equation describes the BFKL Pomeron loop with cut through both Pomerons, and the upper and lower Pomeron couplings given by the meson and quark four-gluon impact factors. The color part of the meson impact factor forbids the coupling of the gluon pair in the color singlet state to the meson, so the Pomerons are formed only between the gluons at the  opposite 
 sides of the unitarity cut. The transverse momentum of the Pomerons in the loop is $\pm \bm{q}$.
In numerical estimates of (\ref{eq:HadronPartonicXsec}) we shall use the approximation of the quark impact factors by the leading Fourier components,  $ \Phi_q\left(y,\bm{k}_i,\bm{q}\right) \to  \Phi_q ^{(0)} \left(y,(\bm{k}_i-\bm{q}/2)^2,q^2\right)$, and
 $\Phi_V\left(\bm{k}_i,\bm{q}\right) \to  \Phi_V ^{(0)} \left((\bm{k}_i-\bm{q}/2)^2,q^2\right)$, cf.\ the discussion performed for the diffractive photoproduction.

Note that we apply in Eq.\ (\ref{eq:HadronPartonicXsec}) the vector meson impact factor $\Phi_{V}$ corresponding to a transition of a transversely polarized gluon to a transversely polarized meson. The gluon is treated within the collinear approximation so it cannot carry the longitudinal polarization.  As in the diffractive photoproduction case the transition of a transversely polarized gluon to a longitudinaly polarized meson is strongly suppressed \cite{Ginzburg:1996vq}.

Special attention should be paid to the target quark and gluon impact factors. At the leading order the parton four-gluon impact factor is a constant function of the gluon momenta, as it is for the two-gluon impact factor.  This follows directly from the point-like nature of the partons. As a result, the four-gluon parton impact factor is proportional to the product of two-gluon impact factors. Next, in our approach we approximate the full Bartels--Kwieci\'{n}ski--Prasza\l{}owicz (BKP) \cite{Bartels:1978fc,Bartels:1980pe,Kwiecinski:1980wb} evolution of the four gluon $t$-channel state in the amplitude squared by the independent evolution of two Pomerons. This approximation is valid in the large $N_c$ limit, as color reconnection between two Pomerons is suppressed by $1/N_c ^2$~\cite{Bartels:1993it}.  Hence, with the factorized form of the parton impact factor, and with the independent Pomeron evolutions, also the evolved four-gluon parton impact factor may be factorized (up to a constant factor) into a product of two-gluon evolved impact factors. Thus, in Eq.\ (\ref{eq:HadronPartonicXsec}) both quark impact factors are evolved independently according to Eq.~(\ref{eq:BFKL1}).

\begin{figure}[t]
\centerline{\includegraphics[width=7cm,angle=0]{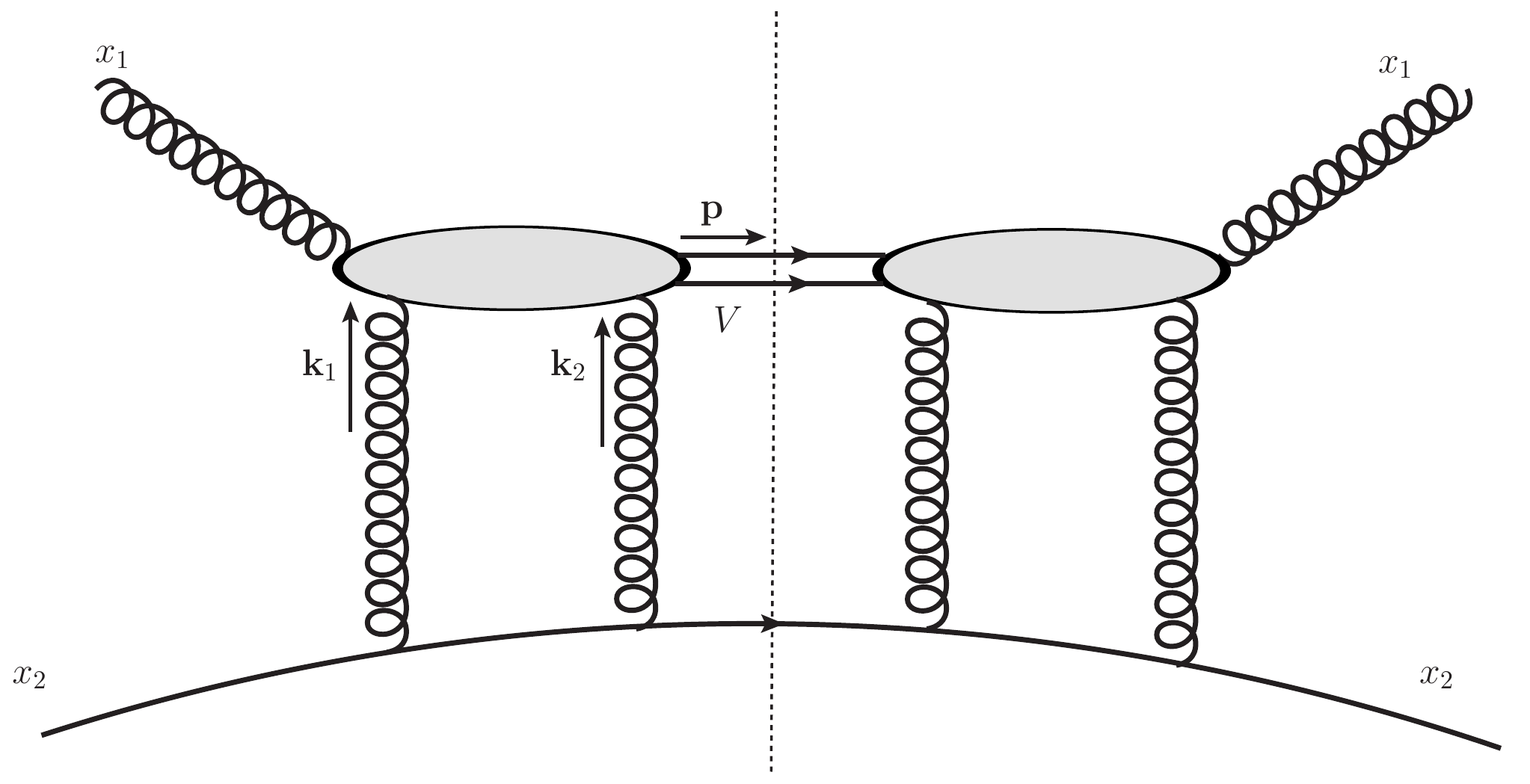}\hspace*{1cm}\includegraphics[width=7cm,angle=0]{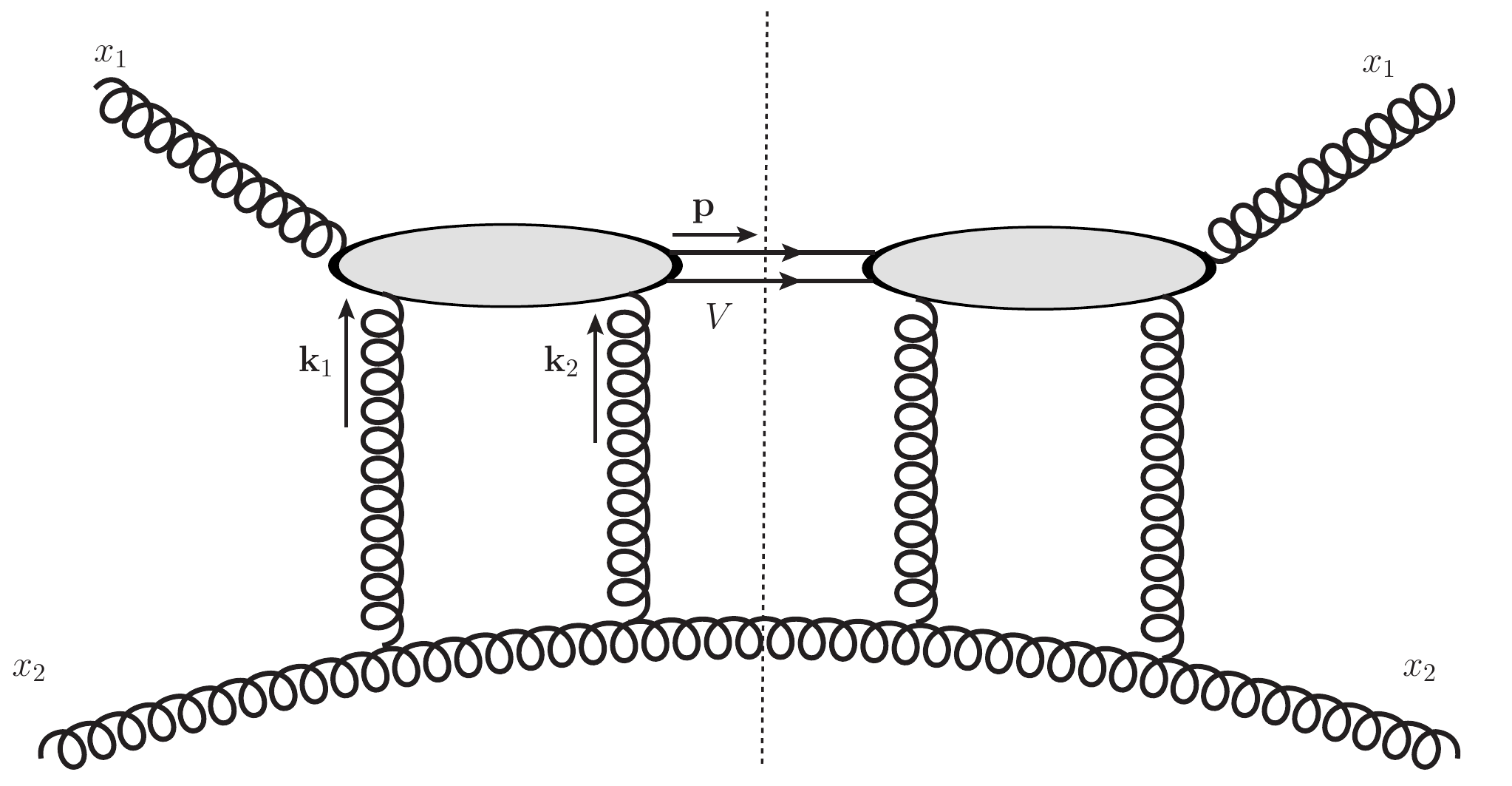}}
\caption{The lowest order diagrams for the amplitude squared for the production of the heavy vector meson through the coupling of $1+2$ gluons off a quark (left) and off a gluon (right).}
\label{fig:borndiagram}
\end{figure}

Let us now discuss the lowest order contribution to the correlated cross section (\ref{eq:HadroXsec1}), obtained by setting $\Phi_q (y,\bm{k}_i,\pm\bm{q}) \to \Phi_{q,0} (\bm{k}_i,\pm\bm{q})$ in Eq.\ \eqref{eq:HadronPartonicXsec}. The corresponding diagrams are depicted in Fig.~\ref{fig:borndiagram}. We see that these diagrams are virtually the same as for the diffractive photoproduction at the lowest order, except for the vector meson vertex, which here contains an incoming gluon instead of a photon. Given the symmetry of the color part of the impact factor, the lowest order $gq \to Vq$ and $gg \to Vg$ cross sections may be obtained from the  $\gamma q \to Vq$ and $\gamma g \to Vg$ cross sections by suitable modifications of the color factors and the coupling constants, while the momentum dependent part remains the same. The explicit result reads
\begin{equation}
\frac{d\sigma_{\mathtt{2-}\mathbb{P}}}{d|\bm{p}|}
 = 32\, \alpha_s^5\, g^2 _{V}\, M_V^2 \frac{1}{|\bm{p}|^7}\,  L^2 _0\left(\frac{M_V^2}{p^2}\right)\, ,
\label{eq:HadronBorn}
\end{equation}
where
\begin{equation}
L_0(u) = 4\pi \frac{1}{u^2-1} \log\frac{\left(1+u\right)^2}{4u}\, .
\end{equation}
This result is a useful benchmark for the cross section (\ref{eq:HadronPartonicXsec}) with the BFKL evolution included.

\subsection{The color factors}

In terms of the BKP equation eigenstates, Eq.\ (\ref{eq:HadronPartonicXsec}) represents exchange of the leading two Pomeron state (where the large $N_c$ limit is implicitly assumed).  
The BKP equation for four gluons, however, allows also for other solutions, like e.g.\ a single Pomeron exchange, in which the elementary gluons are paired into gluon Reggeons. In order to properly project the impact factors on the two Pomeron BKP eigenstates, we apply the following procedure. We start from the Bose symmetry properties of the BKP eigenstates. Given the symmetric kinematic part of the impact factors, it imposes the color symmetry between the pairs of gluons at the same side of the unitarity cut. Also, the color part should be invariant under the interchange of the gluons at the left and right side of the unitarity cut. Employing the relations between invariant tensors of QCD, we find that possible color tensors for the  $t$-channel gluons are
\begin{equation}
P_0 (\{a_n\}) = \delta^{a_1 a_2} \delta^{a_3 a_4},\; 
P_2 (\{a_n\}) = \delta^{a_1 a_3} \delta^{a_2 a_4} + \delta^{a_1 a_4} \delta^{a_2 a_3}, \;
P_d (\{a_n\}) = d^{a_1 a_2 c} d^{a_3 a_4 c},
\end{equation}
where $a_n$, $n=1,\ldots,4$, and $c$ are color indices in the adjoint representation, $a_n$ describe the $t$-channel gluons in the natural order.   The scalar product on the space of the color tensors may be defined as
\begin{equation}
\langle P_A | P_B \rangle = \sum _{\{a_n\} } P_A (\{a_n\}) P_B (\{a_n\}), \;\; A,B=0,2,d.
\end{equation}
It is convenient to use the normalized color tensors
\begin{equation}
\tilde P_A (\{a_n\}) = {P_A (\{a_n\}) \over \sqrt{\langle P_A | P_A \rangle}}\,. 
\end{equation}
The color tensors $\tilde P_0 (\{a_n\})$,  $\tilde P_2 (\{a_n\})$ and  $\tilde P_d (\{a_n\})$ are normalized to one, and orthogonal up to $1/N_c^2$ corrections, $\langle \tilde P_A | \tilde P_B \rangle = \delta_{AB} + {\cal O}(1/N_c^2)$. Since the analysis is performed at the leading order in $N_c$, the tensors $\tilde P_A (\{a_i\})$ may be treated as an othonormal basis.
Therefore  the 
projectors may be defined on the color tensors, corresponding to the BKP eigenstates:
\begin{equation}
{\cal P}_A (\{a_n\},\{b_n\}) = {P_A (\{a_n\})P_A (\{b_n\})
\over  \langle P_A | P_A \rangle} \, .
\end{equation}
In our calculation we project the impact factors on the color state $P_2$ corresponding to the exchange of 
two cut Pomerons.  The color tensors associated with the upper and lower impact factor are denoted by $C_{\alpha} (\{a_n\})$ and $C_{\beta} (\{b_n\})$ correspondingly, and we define the color factor by 
\begin{equation}
C_{\alpha\beta} = \sum_{\{a_n\}} \sum_{\{b_n\}} C_{\alpha} (\{a_n\})  {\cal P}_2 (\{a_n\},\{b_n\})  C_{\beta} (\{b_n\})\, ,
\label{eq:c12}
\end{equation}
where the color projector takes the following form
\begin{equation}
{\cal P}_2 (\{a_n\},\{b_n\}) = {1 \over 2 N_c^2 (N_c^2 -1)} \,
(\delta^{a_1 a_3} \delta^{a_2 a_4} + \delta^{a_1 a_4} \delta^{a_2 a_3} )\,
(\delta^{b_1 b_3} \delta^{b_2 b_4} + \delta^{b_1 b_4} \delta^{b_2 b_3} )\, . 
\end{equation}
The color tensors entering the cross sections are the following:
\begin{equation}
C_{\alpha}(a_1,a_2,a_3,a_4) = {1 \over N_c^2 -1}\sum_c { d^{a_1 a_2 c} d^{a_3 a_4 c} \over 4 N_c^2}\, ,
\label{eq:c1}
\end{equation}
for the color averaged gluon to meson transition amplitude squared, and
\begin{equation}
C_{\beta}(b_1,b_2,b_3,b_4) = {1 \over \mathrm{dim} ( R_i ) }\, \mathrm{Tr} \left(\hat{T}_{R_i} ^{b_4} \hat{T}_{R_i} ^{b_3} \hat{T}_{R_i} ^{b_2} \hat{T}_{R_i} ^{b_1}\right) \; ,
\label{eq:c2}
\end{equation}
for the color averaged scattering amplitude squared of the parton in color representation $R_i$.
The explicit form of color factors $C_{\alpha\beta}$ follows from Eqs.\ (\ref{eq:c12}), (\ref{eq:c1}), (\ref{eq:c2}) 
 and reads 
\begin{equation}
C_{gq} = \frac{\left(N_c^2-4\right)\left(N_c^2-2\right)}{16N_c^7}\, , \qquad
C_{gg} = \frac{3}{8}\, \frac{N_c^2-4}{\left(N_c^2-1\right)N_c^3} \, .
\label{eq:HadroColorfact}
\end{equation}

\subsection{Vector meson hadroproduction in conformal representation of the BFKL Pomeron}

In what follows we shall recall the Lipatov solution to the non-forward leading logarithmic BFKL equation by means of the conformal eigenfunctions \cite{Lipatov:1985uk}, and the application to diffractive scattering. Then we shall employ the formalism to describe the vector meson hadroproduction in the two Pomeron exchange mechanism. The solution to the non-forward BFKL evolution equation can be presented in the momentum or in the coordinate space. 
To this end, for a generic 2-dimensional transverse momentum $\kb=(k_x,k_y)$ we introduce the complexified momenta $(k,\bar{k})$, where
\be
k=k_x+i k_y, \qquad \bar{k} = k_x - i k_y \, .
\ee
Similarly, for the transverse coordinate space, the 2-dimensional coordinates $\bm{\rho}=(\rho_x,\rho_y)$ will be traded for the complex variables
\be
\rho=\rho_x+i \rho_y, \qquad  \bar\rho=\rho_x - i \rho_y \,. 
\ee
In what follows we set $s_0 = 0$ and use the original form of the non-forward BFKL equation. 
It was shown by Lev Lipatov \cite{Lipatov:1985uk, Lipatov:1996ts} that the BFKL equation in the leading logarithmic approximation is invariant under
 the conformal transformations of the complexified transverse positions of the Reggeized gluons,
\be
\rho \rightarrow \frac{a \rho + b}{c \rho +d} \;,
\ee
for arbitrary complex parameters $a,b,c,d$.
In analyses of high energy scattering amplitudes ${\cal A}(s,t)$ it is customary to use 
 the 
Mellin moments $\omega$ conjugate to $s$ \cite{Lipatov:1996ts},
\be
{\cal A}(s,t) = |s| \int {d\omega \over 2\pi i}  A(\omega,t) s^{\omega}.
\ee
The solution for the gluon Green's function is then represented as
\begin{multline}
G_{\omega}(\rho_1,\rho_2;\rho_1',\rho_2') = \sum_{n=-\infty}^{+\infty} \int_{-\infty}^{+\infty} d\nu \frac{\nu^2+n^2/4}{(\nu^2+(n-1)^2/4)(\nu^2+(n+1)^2/4)} \\
\times\;\int d^2 \rho_0 \frac{E_{n\nu}(\rho_{10},\rho_{20}) E^*_{n\nu}(\rho_{1'0},\rho_{2'0})}{\omega-\omega_n(\nu)} \,,
\end{multline}
where $\omega_{n}(\nu)$ is the 
 LL BFKL  eigenvalue
\be
\omega_n(\nu) \;  = \; \frac{N_c \alpha_s}{\pi} 
\left[ 2 \psi(1)-\psi\left(\frac{|n|}{2}+\frac{1}{2}+i\nu\right)-\psi\left(\frac{|n|}{2}+\frac{1}{2}-i\nu\right) \right]\;,
\label{eq:lleigen}
\ee
and $\psi$ are polygamma functions, the 
functions $E_{n,\nu}$ are conformal eigenfunctions defined as
\be
E_{n\nu}(\rho_1,\rho_2) \; = \; \bigg( \frac{\rho_{12}}{\rho_1\rho_2}\bigg)^h\, \bigg( \frac{\rho^*_{12}}{\rho^*_1\rho^*_2} \bigg)^{\tilde{h}}\, ,
\label{eq:confeigen}
\ee
where the powers $h,\tilde{h}$ are the conformal weights 
\be
h=\frac{1}{2}+\frac{n}{2}+i\nu,\;\;\;\;\;\;\tilde{h}=\frac{1}{2}-\frac{n}{2}+i\nu \, .
\ee

The above form of the BFKL Green's function may be Fourier transformed to the transverse momentum representation, and inverse-Mellin-transformed in $\omega$ to the rapidity space
\begin{multline}
{\cal G}_y(\bm{k}_1,\bm{k}' _1, \bm{q}) =  {1\over (2\pi)^6}\sum_{n=-\infty}^{+\infty} \int_{-\infty}^{+\infty} 
d\nu\,\frac{\nu^2+n^2/4}{(\nu^2+(n-1)^2/4)(\nu^2+(n+1)^2/4)} \\
\times\; \exp (\omega_n(\nu)y)\, \tilde E_{n\nu}(\bm{k}_1,\bm{q}) \, \tilde E^*_{n\nu}(\bm{k}_1',\bm{q}) \,,
\end{multline}
where
\be
\tilde E_{n\nu}(\bm{k}_1,\bm{q}) = \int d^2 \bm{\rho}_1  d^2 \bm{\rho}_2 \, \exp(i\bm{k}_1 \bm{\rho}_1 + i (\bm{q}-\bm{k})\bm{\rho}_2)\; ,
\ee
are the conformal eigenfunctions in the momentum representation. 
The analytic form of the eigenfunctions $E_{n\nu}(\bm{k}_1,\bm{q})$ was derived in \cite{Navelet:1997xn}.
 It is rather lengthy, so instead of listing it here we refer the reader to the original paper.
In the limit $y \to 0$, one finds 
\be
{\cal G}_y(\bm{k}_1,\bm{k}' _1, \bm{q})  \to {\delta^2(\bm{k}_1-\bm{k}'_1) \over \bm{k}_1 ^2 (\bm{q} - \bm{k}_1)^2} \, . 
\ee

The dominant imaginary part of the amplitude for the BFKL Pomeron exchange between leading order impact factors $\Phi_{A,0}(\bm{k}_1,\bm{q})$ and  $\Phi_{B,0}(\bm{k}'_1,\bm{q})$ reads
\begin{equation}
\mathrm{Im}\,  \mathcal{A}\left(\hat s,t=- q^2 \right)
 = \hat s \int\! \frac{d^2\bm{k}_1 d^2\bm{k}'_1}{2\pi} \,
\Phi_{A,0}\left(\bm{k}_1,\bm{q}\right)\,
{\cal G}_y(\bm{k}_1,\bm{k}' _1, \bm{q})
\Phi_{B,0}\left(\bm{k}'_1,\bm{q}\right).
\label{eq:bfklgreens} 
\end{equation}
It is convenient to define the projection of the impact factors on the conformal eigenfunctions
\be
I^A _{n,\nu}(\bm{q}) = \int {d^2 \bm{k}_1 \over (2\pi)^2 }  \Phi_{A,0}(\bm{k}_1,\bm{q})\,  \tilde E_{n\nu}(\bm{k}_1,\bm{q}),
\ee
and analogously for the index $B$. This leads to the following form of the Pomeron exchange amplitude
\begin{multline}
\mathrm{Im}\,  \mathcal{A}\left(\hat s,t=- q^2 \right)
 = \hat s   {1\over (2\pi)^3}\sum_{n=-\infty}^{+\infty} \int_{-\infty}^{+\infty} 
d\nu\,\frac{\nu^2+n^2/4}{(\nu^2+(n-1)^2/4)(\nu^2+(n+1)^2/4)} \\
\times\; \exp (\omega_n(\nu)y)\, I^A _{n,\nu}(\bm{q}) \, [I^B_{n,\nu}(\bm{q})]^* \,.
\end{multline}
The impact factors $I^q _{n,\nu}(\bm{q})$ and $I^V_{n,\nu}(\bm{q})$ were computed in the analytic form in Refs.\ \cite{Mueller:1992pe,Bartels:1996fs,Motyka:2001zh,Enberg:2002zy}. In the calculation of vector meson hadroproduction we shall need the quark impact factor, which is treated within the Mueller--Tang scheme \cite{Mueller:1992pe,Motyka:2001zh}. It takes the form
\be
I^q _{n,\nu} = -{4\pi \alpha_s i^n \over q} \left( {q^2 \over 4} \right) ^{i\nu} e^{-in\phi_q}
{\Gamma(1/2 + n/2 - i\nu) \over \Gamma(1/2 + n/2 + i\nu)}\, ,
\label{eq:quarkMT}
\ee
where $\phi_q$ is the polar angle of the vector $\bm{q}$ in the transverse plane.

The formalism may be also applied to the two-Pomeron exchange process, assuming independent BFKL evolution of the Pomerons.
Hence, we rewrite Eq.\ (\ref{eq:HadronPartonicXsec}) as 

\begin{multline}
{d\sigma_{\mathtt{2-}\mathbb{P}}\left(x_1 x_2 S,t,\mu \right) \over d^2\bm{p}}
= {\alpha_s \over 16 \pi^2 \alpha_{em} e_q^2} 
\int\!\frac{d^2\bm{k}_1 d^2\bm{k}'_1}{2\pi} \int\!\frac{d^2\bm{k}_2 d^2\bm{k}'_2}{2\pi}
  \int\!d^2\bm{q} \,\, \delta^{2}\left(\bm{k}_1+\bm{k}_2-\bm{p}\right) \\
\times \,
\Phi_V\left(\bm{k}_1,\bm{k}_1+\bm{k}_2\right)
\Phi_V\left(\bm{q}-\bm{k}_1,-\bm{k}_1-\bm{k}_2\right)
\\
\times \, 
{\cal G}_y(\bm{k}_1,\bm{k}' _1, \bm{q}) \,{\cal G}_y(\bm{k}_2,\bm{k}' _2, -\bm{q}) \,
\Phi_{q,0}\left(\bm{k}' _1 ,\bm{q}\right)\Phi_{q,0}\left(\bm{k}' _2,-\bm{q}\right)
\, .
\label{eq:conformal2P}
\end{multline}
Using the representation of the BFKL Green's functions by the conformal eigenfunctions in the momentum space this may be
rewritten as,
\begin{multline}
{d\sigma_{\mathtt{2-}\mathbb{P}}\left(x_1 x_2 S,t,\mu \right) \over d^2\bm{p}}
= {\alpha_s \over 16\pi^2 \alpha_{em} e_q^2} 
  \int\!d^2\bm{q} \sum_{n_1=-\infty}^{+\infty} \int_{-\infty}^{+\infty} 
{d\nu_1 \over (2\pi)^3}\,\frac{\nu_1^2+n_1^2/4}{(\nu_1^2+(n_1-1)^2/4)(\nu_1^2+(n_1+1)^2/4)} \\
\times
\sum_{n_2=-\infty}^{+\infty} \int_{-\infty}^{+\infty} 
{d\nu_2 \over (2\pi)^3}\,\frac{\nu_2^2+n_2^2/4}{(\nu_2^2+(n_2-1)^2/4)(\nu_2^2+(n_2+1)^2/4)} \, 
\exp[\bar\alpha_s y (\omega_{n_1}(\nu_1) + \omega_{n_2}(\nu_2))]\\ 
\times
I^{V\otimes V} _{n_1,\nu_1;n_2,\nu_2}(\bm{p},\bm{q}) \,\left[I^{q}_{n_1,\nu_1}(\bm{q})\, I^{q}_{n_2,\nu_2}(-\bm{q})\right]^*
\, .
\label{eq:conformal2Pb}
\end{multline}
A new non-trivial object that appears in this equation, $I^{V\otimes V} _{n_1,\nu_1;n_2,\nu_2}(\bm{p},\bm{q})$ is a projection of a pair of the vector meson impact factors (coming from the amplitude and its complex conjugate) on the conformal eigenfunctions:
\begin{multline}
I^{V\otimes V} _{n_1,\nu_1;n_2,\nu_2}(\bm{p},\bm{q}) 
=  \int\!\frac{d^2\bm{k}_1}{(2\pi)^2} \int\!\frac{d^2\bm{k}_2}{(2\pi)^2}
\delta^{2}\left(\bm{k}_1+\bm{k}_2-\bm{p}\right)\, 
\tilde E_{n_1,\nu_1}(\bm{k}_1,\bm{q})\, \tilde E_{n_2,\nu_2}(\bm{k}_2,-\bm{q}) \\
\times \, \Phi_V\left(\bm{k}_1,\bm{p}\right)
\Phi_V\left(\bm{q}-\bm{k}_1,\bm{p}\right). 
\label{eq:VVprojection}
\end{multline}
Note that in contrast to the diffractive scattering, in this expression the vector meson impact factors  $\Phi_V$ are coupled to two different conformal eigenfunctions each. In other words, the two meson impact factors and the two conformal eigenfunctions are all entangled and integral (\ref{eq:VVprojection}) cannot be factorized. 

In order to cross check the numerical predictions  for the vector meson hadroproduction  by
 the  two Pomeron exchange, obtained primarily by solving the Eq.~(\ref{eq:BFKL1}) and Monte Carlo integration in Eq.~(\ref{eq:HadroXsec1}),  we calculated numerically integral (\ref{eq:VVprojection}).
  Then, the partonic cross section in the conformal representation was evaluated using Eq.\ (\ref{eq:conformal2Pb}). In this approach a good convergence of the emerging numerical sums and integrals was found for $\bar\alpha_s y \gtrapprox 1$, where the results were found to coincide with  results obtained within the direct approach described in Section\ \ref{sec:HadroDirect}.

\section{Properties of the Pomeron loop at the parton level}
\label{sec:PomLoopParton}

In this subsection we  analyze the properties of the partonic cross section which includes Pomeron loop contribution \eqref{eq:HadronPartonicXsec}. In all calculations we shall set the infrared cutoff $s_0$  to be equal to $0.5 \, {\rm GeV^2}$.

\begin{figure}[t]
	\centerline{\includegraphics[width=0.7\columnwidth,angle=0]{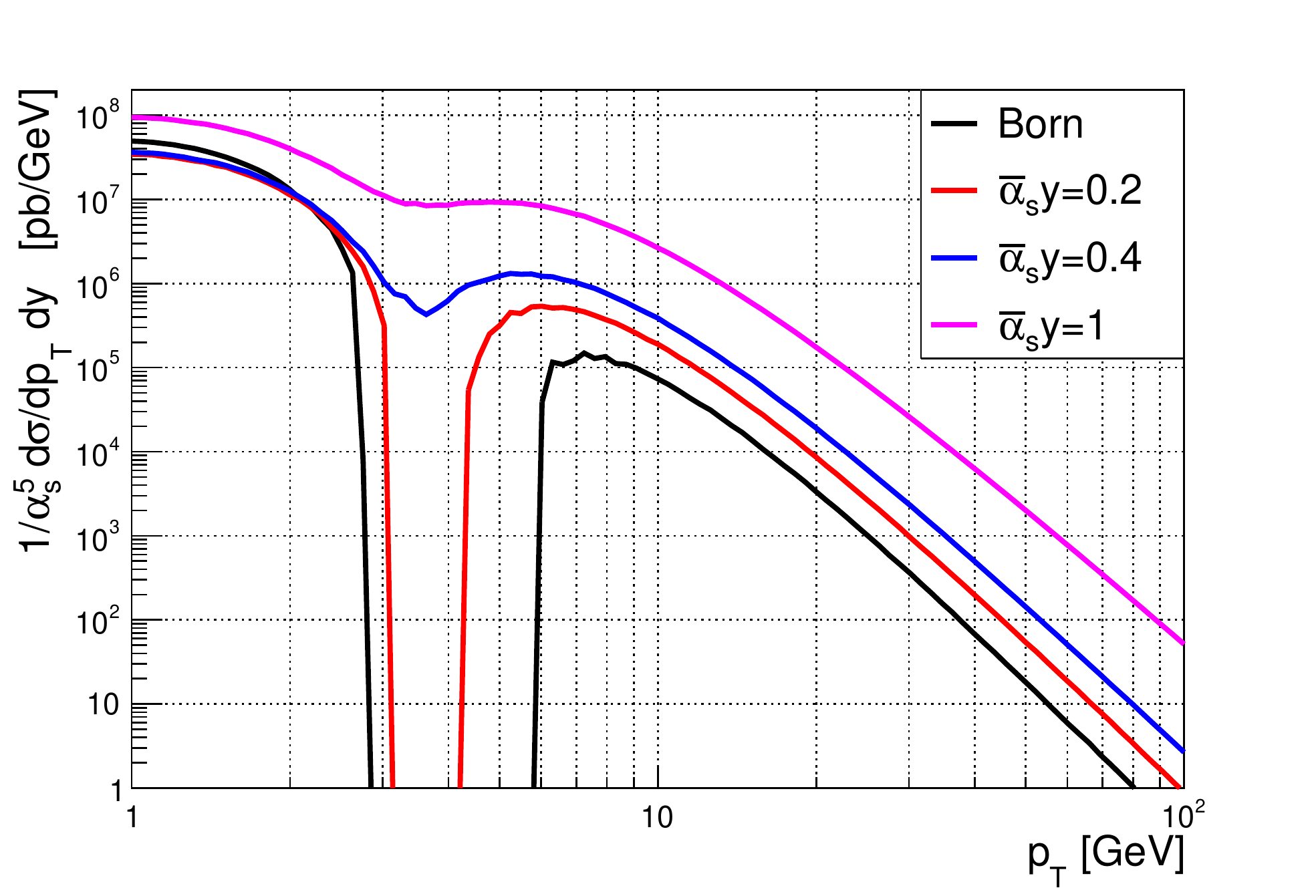}}
	\caption{Partonic cross section with the correlated Pomeron loop, Eq.~\eqref{eq:HadronPartonicXsec} as a function of transverse momentum for different values of rapidity. }
	\label{fig:PartXsecdpt}
\end{figure}

In Fig.~\ref{fig:PartXsecdpt} the result of calculation of the partonic cross section from Eq.~\eqref{eq:HadronPartonicXsec}
is shown as a function of the transverse momentum $p_T$ of the produced vector meson for fixed values of rapidity.  The normalization is given by Eq.~\eqref{eq:HadronPartonicXsec}, but the cross section is divided by $\alpha_s^5$. The reason to divide out the strong coupling constant is that it introduces additional transverse momentum dependence, and we want to illustrate the transverse momentum dependence originating purely from the kinematical parts of the cross section and the BFKL Pomeron exchange.
We observe that the cross section has a characteristic dip at values of transverse momentum around the vector meson mass $p_T \sim M_V$. The dip appears because of a change of the sign of the partonic amplitude. Its appearance and evolution is well known from analyses of the diffractive vector meson production at large momentum transfer \cite{Enberg:2002zy,Enberg:2003jw,Poludniowski:2003yk}. The dip appears at the lowest order for which the kinematical parts of the amplitudes for diffractive photoproduction and the triple gluon fusion hadroproduction are the same. For the diffractive photoproduction  the dip moves to larger values of $p_T$ with increasing rapidity. In the hadroproduction where the BFKL evolution acts in a different way, the dip becomes shallower with increasing rapidity length of the BFKL evolution. This effect is especially visible at the highest plotted value of  the LL evolution variable $\bar{\alpha}_s y=1$. 
 The fall-off with the transverse momentum is  $1/p_T^n$ with power to be equal approximately $n = 5\div6$. 
In the next figure, Fig.~\ref{fig:PartXsecdy} the result of the 
same calculation as a function of rapidity length of the evolved Pomerons is shown for fixed values of $p_T$. The exponential growth with rapidity is clearly visible with the exponent being equal to approximately $\sim 0.5$ which is consistent with the exchange of two LL Pomerons.

\begin{figure}[t]
	\centerline{
		\includegraphics[width=0.7\columnwidth,angle=0]{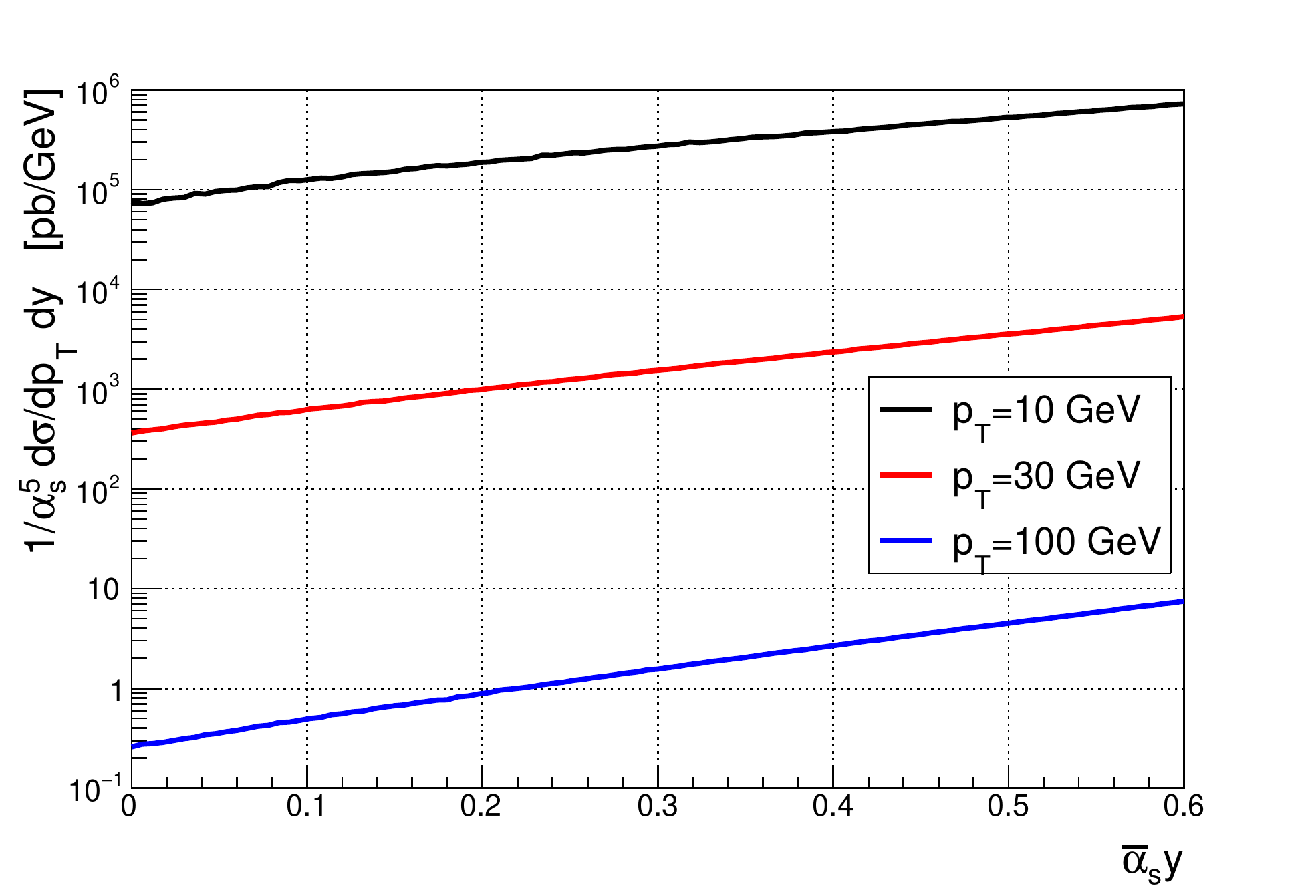}}
	\caption{Partonic cross section with the correlated Pomeron loop, Eq.~\eqref{eq:HadronPartonicXsec} as a function of rapidity for different values of  transverse momentum.}
	\label{fig:PartXsecdy}
\end{figure}

\begin{figure}[t]
	\centerline{\includegraphics[width=0.7\columnwidth,angle=0]{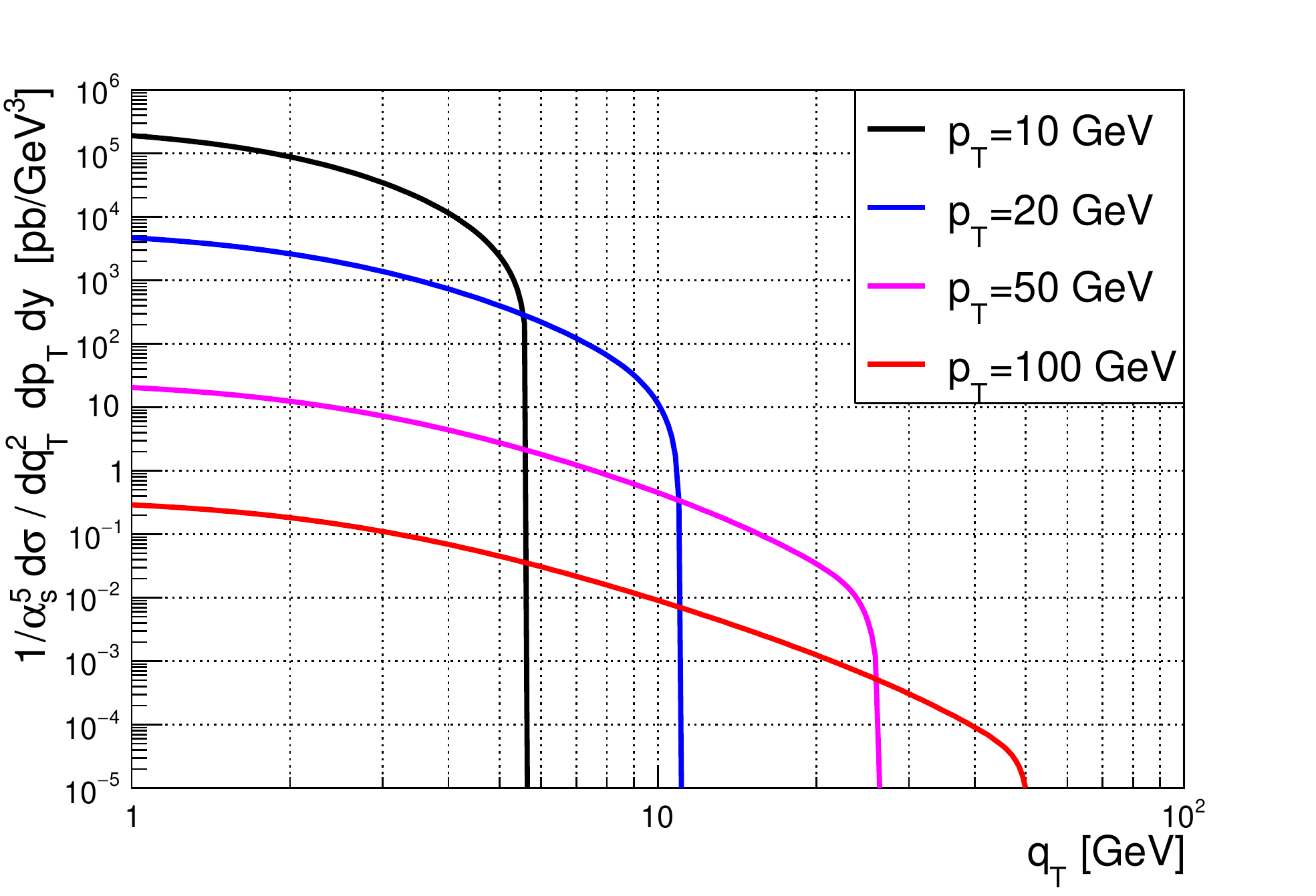}}
	\caption{Partonic cross section with the correlated Pomeron loop, Eq.~\eqref{eq:HadronPartonicXsec} unintegrated over the momentum transfer $q_T$  as a function of the  momentum transfer for different values of the transverse momentum of the produced vector meson.  Evolution variable  in the BFKL equation is $\bar{\alpha}_s y=0.4$.  }
	\label{fig:PartXsecdqtdpt}
\end{figure}

\begin{figure}[t]
	\centerline{
		\includegraphics[width=0.7\columnwidth,angle=0]{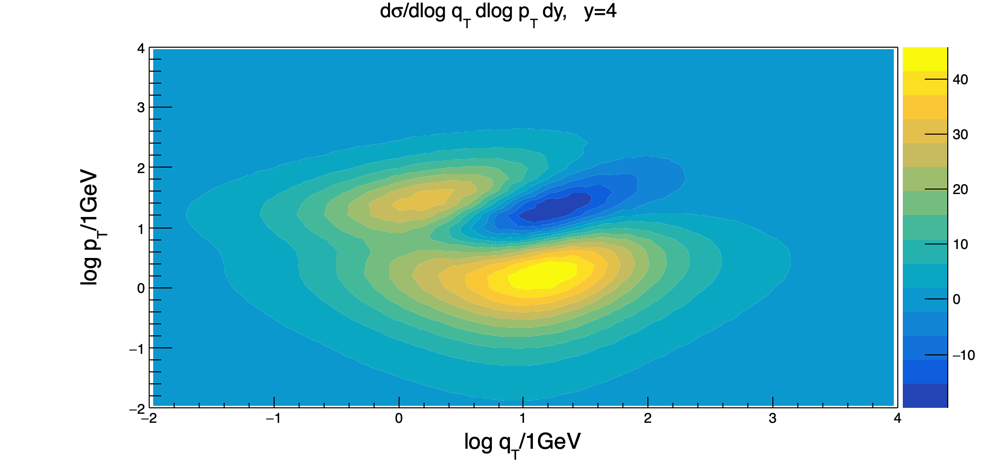}}
	\caption{Partonic cross section with the correlated Pomeron loop, Eq.~\eqref{eq:HadronPartonicXsec} as a function of the momentum transfer $q_T$ and the transverse momentum of the produced vector meson $p_T$. Evolution variable  in the BFKL equation is $\bar{\alpha}_s y=0.4$.  }
	\label{fig:PartXsec2D}
\end{figure}

In order to better understand the properties of the Pomeron loop we also present calculations in which the integration of the momentum transfer $q_T$ is not performed. The results for $\frac{d\sigma}{d q_T^2 dp_T dy}$ are shown in Figs.~\ref{fig:PartXsecdqtdpt} and \ref{fig:PartXsec2D}. In the first figure Fig.~\ref{fig:PartXsecdqtdpt} the partonic cross section is shown as a function of the momentum transfer $q_T$ for different values of the transverse momentum $p_T$. Rapidity in the BFKL evolution is fixed to be equal to $y=4$. We observe that the distribution exhibits a sharp cutoff at values of the  momentum transfer which are approximately half of the value of the transverse momentum $q_T \sim p_T /2$. This cutoff stems from the coupling of the Pomerons to the vector meson impact factor. The value of the integrand becomes slightly negative above the cutoff. This is best visible in the 2-dimensional projection of $\frac{d \sigma}{d\log p_T d\log q_T dy}$ onto $(\log q_T,\log p_T)$ plane, where we observe the cutoff in the form of the dip in the integrand of the cross section in these variables. Note that in the hadroproduction cross section the momentum transfer $q_T$ through the Pomerons in not an observable, and the negative values in some regions of  $\frac{d \sigma}{d\log p_T d\log q_T dy}$ do not imply negativity of any observed differential cross sections.
 Of course, when integrated over the $q_T$ the cross section remains positive definite as it should be.



\section{Numerical results}
\label{sec:Results}
In this section we shall discuss the phenomenological implications of the correlated double-Pomeron exchange contribution to $J/\psi$ and $\Upsilon$ hadroproduction. First, however, we will perform a numerical analysis of the diffractive photoproduction case described in Section~\ref{sec:Diff} in the context of the most recent HERA data \citep{Aktas:2003zi}. We treat this as a benchmark calculation and an opportunity to fix the (very few) theory parameters.

Our numerical calculations are performed in two ways. First, we use the Monte Carlo implementation of the cross section formulae which allows for a great flexibility in choice of distributions one can study. In particular it allows for  usage of consistent observables with those measured. Second, we use  direct implementations of the integrals, which serves as a cross-check and a useful tool for fast studies of smooth distributions. 

The analysis of diffractive $J/\psi$ photoproduction at $\sqrt{S}=318\,\mathrm{GeV}$ \citep{Aktas:2003zi} was performed by H1 collaboration and was found to be consistent with the previous measurements by ZEUS \citep{Chekanov:2002rm}. The latter was addressed in the literature at length in the BFKL context,  see  \citep{Forshaw:2001pf,Enberg:2002zy}. The newer data \citep{Aktas:2003zi} cover larger $\gamma$--$p$ c.m.s.\ energy range $50<\sqrt{s}<200\,\mathrm{GeV}$ (vs $80<\sqrt{s}<120\,\mathrm{GeV}$ at ZEUS) and much higher momentum transfers: $2<|t|<30\,\mathrm{GeV}^2$ (vs $|t|<12\,\mathrm{GeV}^2$ at ZEUS).
We shall be interested in the cross section as a function of $t$. Since we have the Monte Carlo implementation of the hadronic cross section,  we can easily apply the kinematic cuts to be very close to the experimental ones. In particular, instead of calculating the cross section at fixed average $\gamma$--$p$ c.m.s.\ energy as it was the case for previous studies \citep{Forshaw:2001pf,Enberg:2002zy}, we can generate explicitly the equivalent photon flux \citep{Budnev:1974de}
\begin{equation}
f_{\gamma}\left(z,Q_{\mathrm{max}}\right) = \frac{\alpha}{2\pi}\, 
\frac{1+(1-z)^2}{z}\,\log\left(\frac{(1-z)Q_{\mathrm{max}}^2}{z m_e^2}\right)\, ,
\label{eq:photonflux}
\end{equation}
 and calculate the weighted $\gamma$--$p$ cross section. Above, $m_e$ is the electron mass and $Q_{\mathrm{max}}\sim M_V$.  Following \citep{Aktas:2003zi}, we also restrict the mass of the diffractively produced state via the relation
\begin{equation}
M_Y^2<\frac{|t|}{x}\, ,
\end{equation}
with $M_Y = 30\,\mathrm{GeV}$. 

\begin{figure}[t]
\centerline{\includegraphics[width=10cm,angle=0]{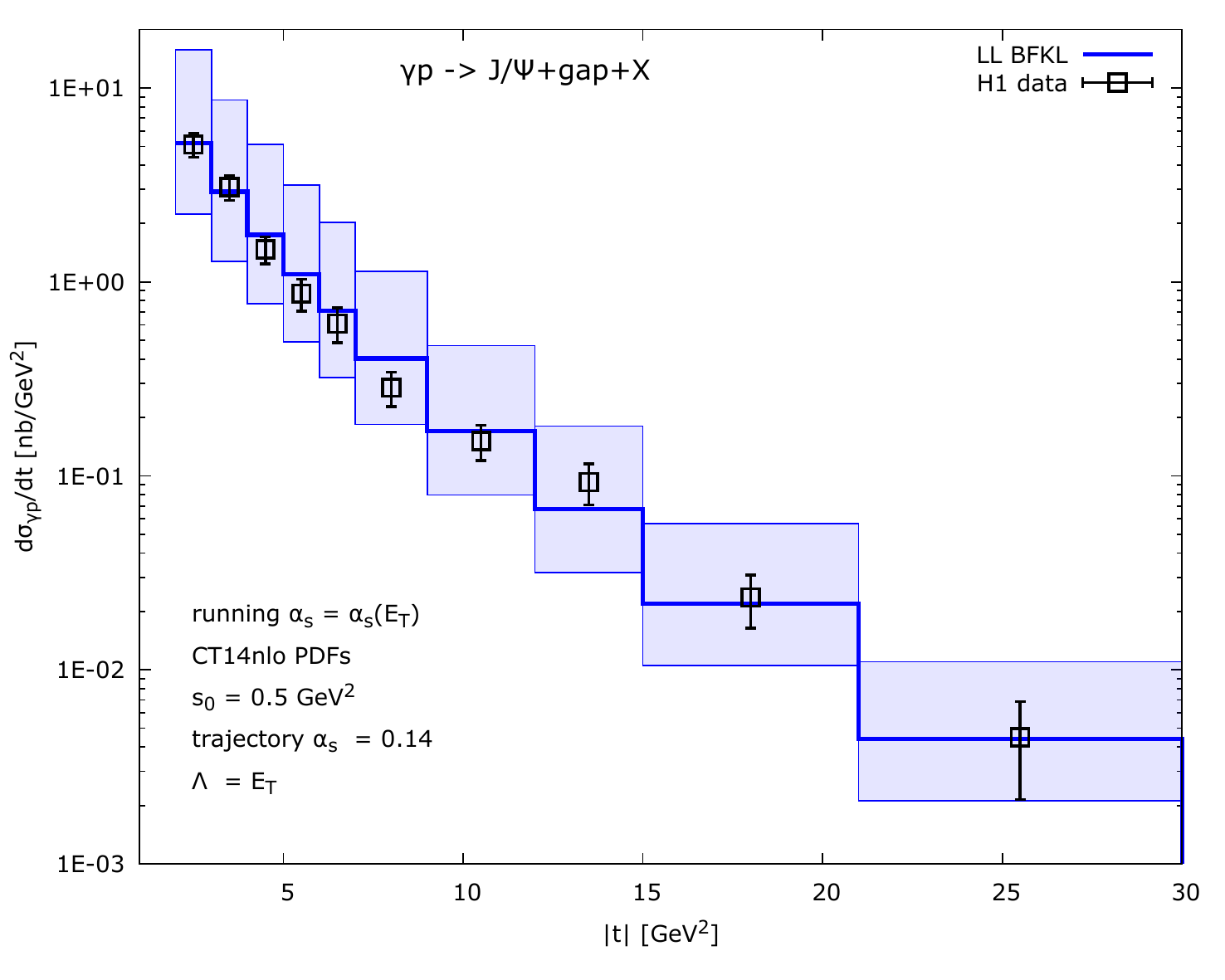}}
\caption{The differential cross section $d\sigma/dt$ for the diffractive $J/\psi$ production vs H1 data \citep{Aktas:2003zi}  for the coupling constant entering the LL BFKL trajectory $\bar\alpha_s=0.14$.}
\label{fig:result_diffraction1}
\end{figure}

\begin{figure}[t]
\centerline{\includegraphics[width=10cm,angle=0]{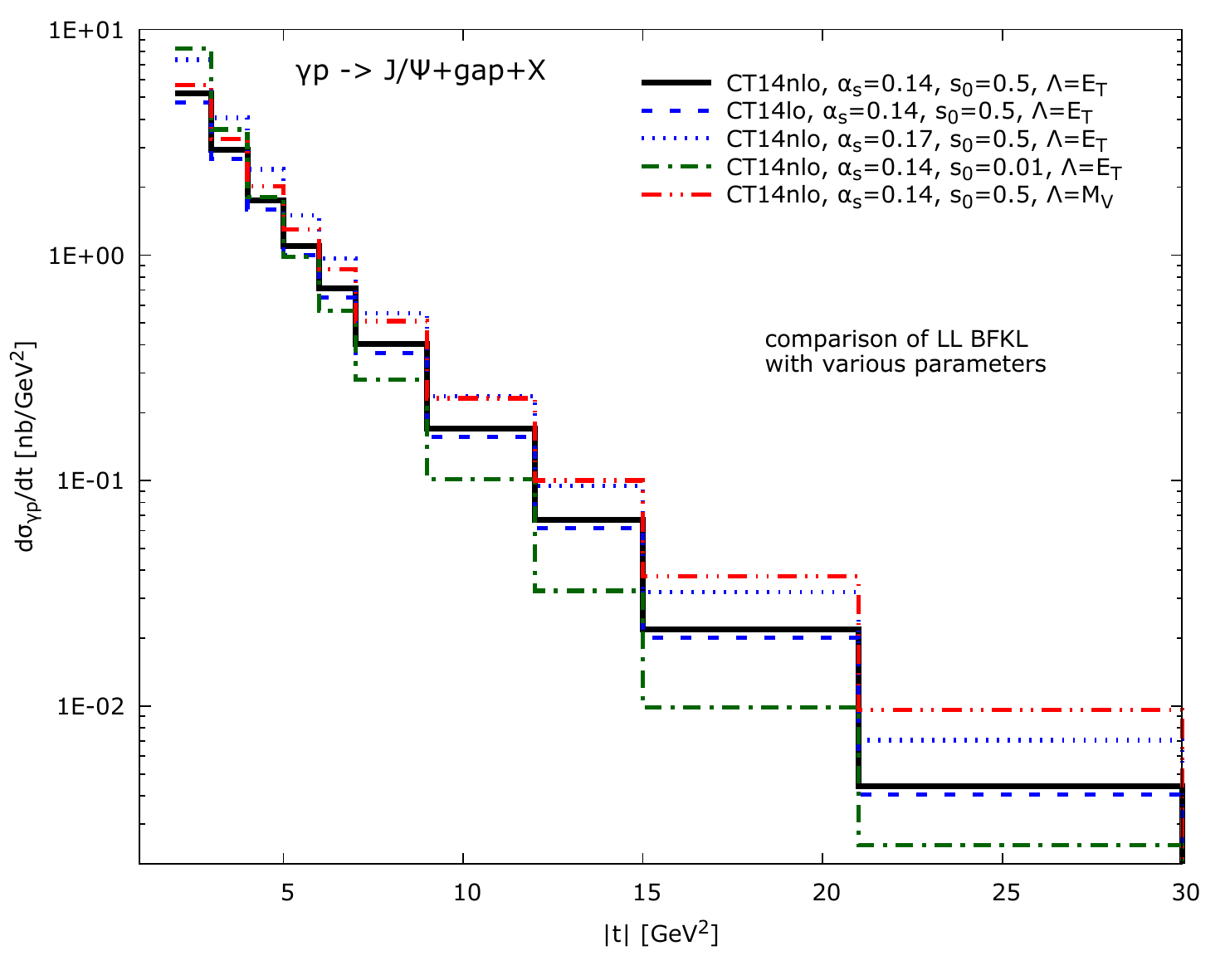}}
\caption{Comparison of the differential cross sections $d\sigma/dt$ for the diffractive $J/\psi$ production for some configurations of the model parameters.}
\label{fig:result_diffraction2}
\end{figure}

There are two main model parameters: the infrared cutoff $s_0$ and the slope of the trajectory in the evolution equation (\ref{eq:BFKL1}) given by the fixed strong coupling constant. The remaining freedom concerns: (i) the $\Lambda$ parameter in Eq.~(\ref{eq:rap1}) which we set to $E_T=\sqrt{p_T^2+M_V^2}$, (ii) the choice of the hard scale $\mu$ entering the strong coupling constant and the PDFs, (iii) the choice of the PDF set itself. Starting with the last point, we use the CT14nlo set \cite{Lai:2010vv}. 
The choice of the hard scale is, in principle, more subtle 
as $\alpha_s$ may be running with a different scale in different components of the calculation. However, we simply set $\mu^2 = p_T^2 + M_V^2$ as the scale for the coupling constant in the impact factors 
and vary the scale by a factor of two to estimate the theoretical uncertainty. 
In the LL BFKL kernel the fixed effective value of the strong coupling constant $\bar\alpha_s=0.14$ is used. This value was adjusted to get a good description of the diffractive photoproduction data. It effectively parametrizes the effect of higher order corrections to the BFKL evolution. The results of calculations in this setup are presented in Fig.~\ref{fig:result_diffraction1} together with H1 data. 
The sensitivity to the scale variation is visualized by the shaded boxes. We observe excellent agreement with data for the assumed parameters. In Fig.~\ref{fig:result_diffraction2} we also show in one plot how different components of the model influence the results.  In this study, the most drastic changes are generated either by decreasing the $s_0$ cutoff to $0.01\,\mathrm{GeV}^2$ or by increasing the value of the intercept. Although in both cases the cross section gets increased significantly by a similar amount, the large $|t|$  tail is different. The change of the $\Lambda$ scale to the constant value equal to $M_V$ significantly hardens the spectrum. The change to the leading order PDFs  has negligible effect.

\begin{figure}[t]
\centerline{\includegraphics[width=8cm,angle=0]{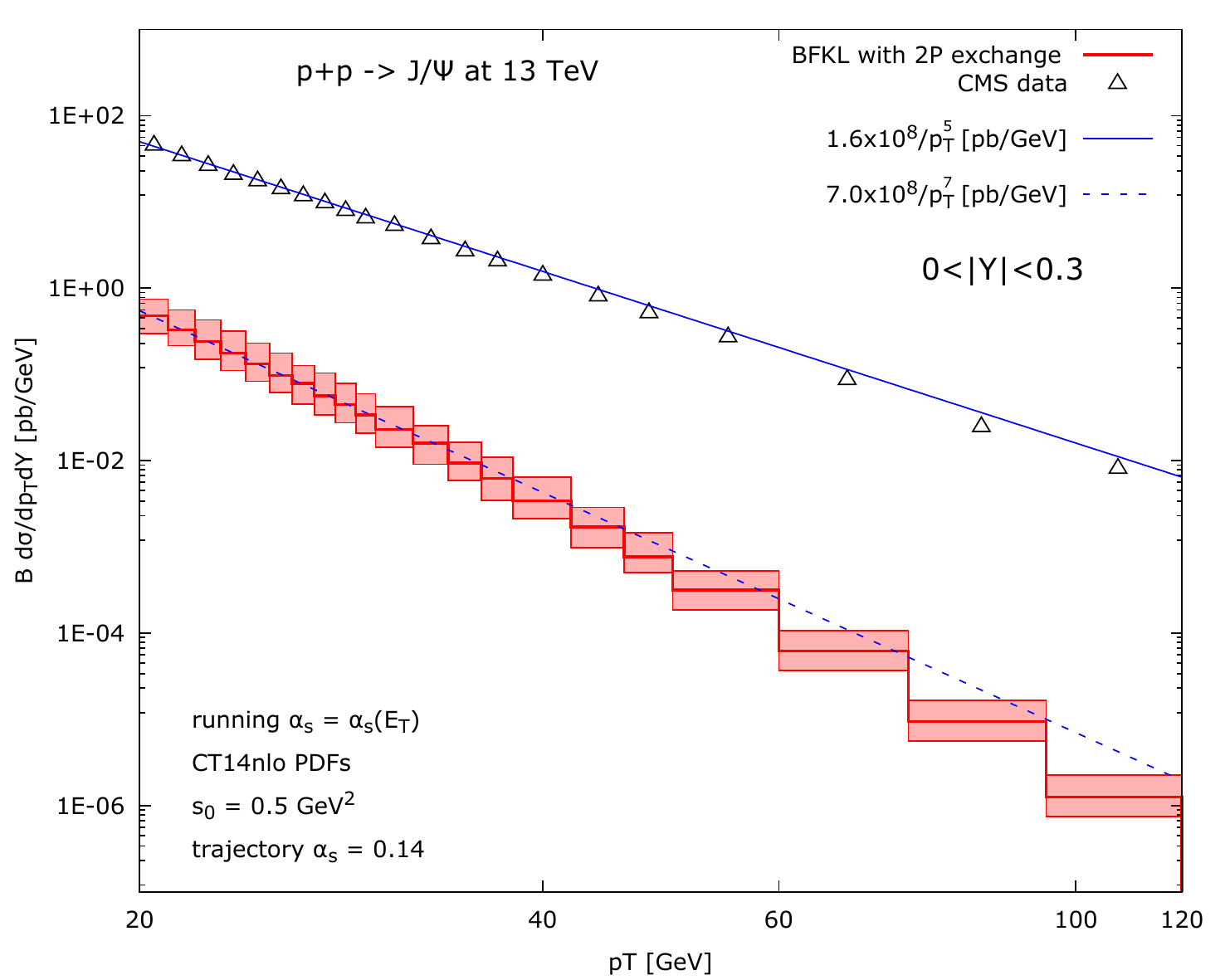}\quad
\includegraphics[width=8cm,angle=0]{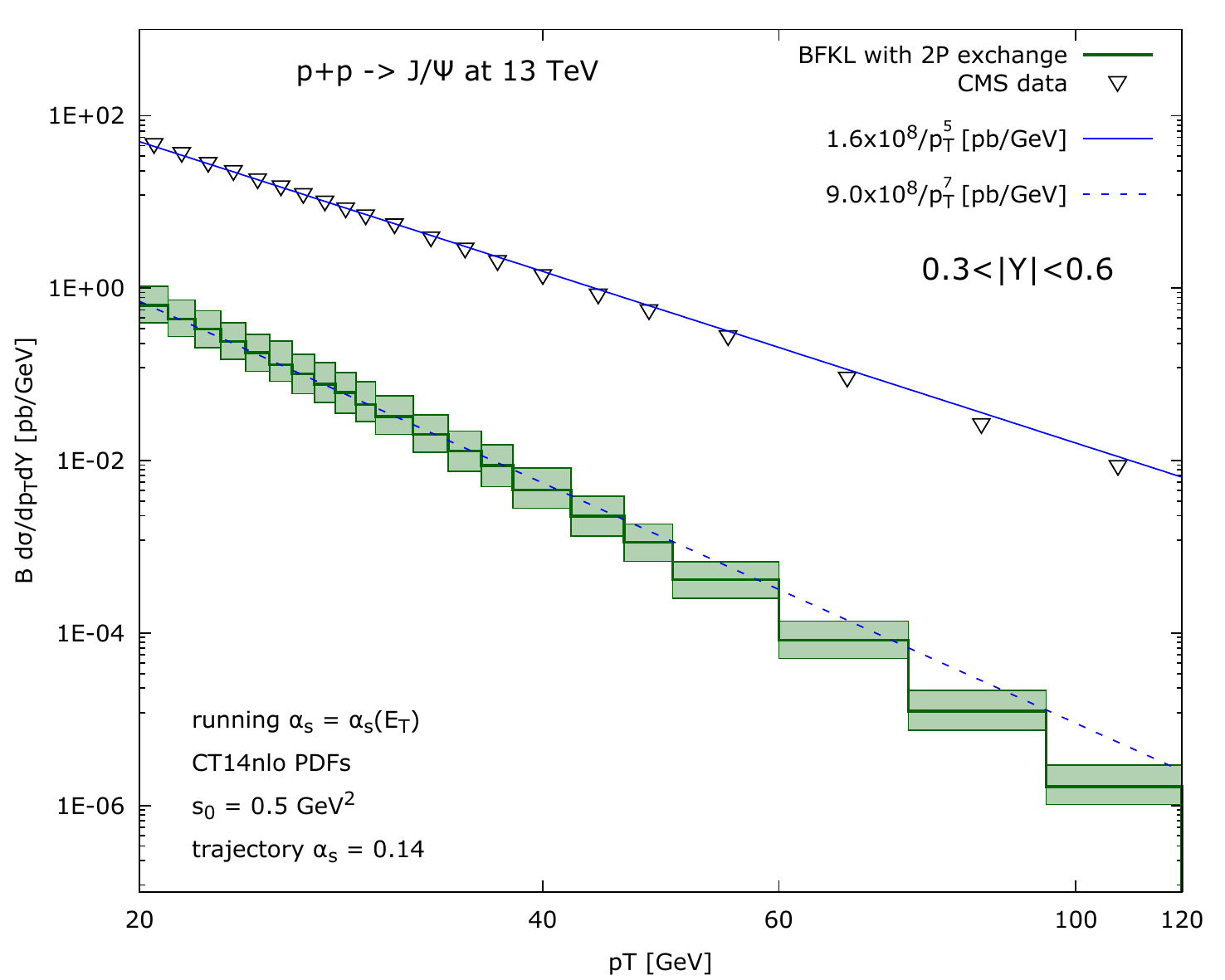}}

\centerline{\includegraphics[width=8cm,angle=0]{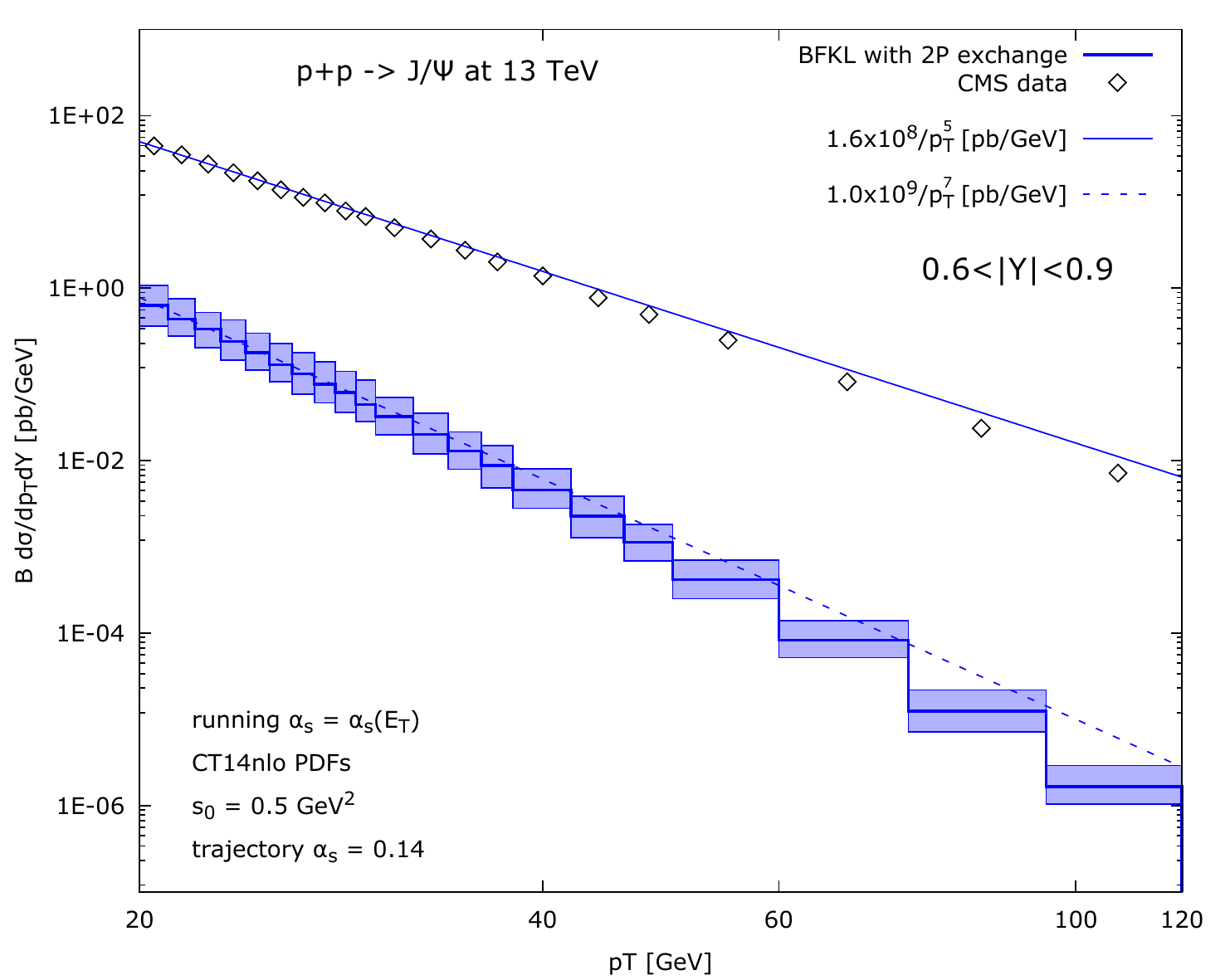}\quad
\includegraphics[width=8cm,angle=0]{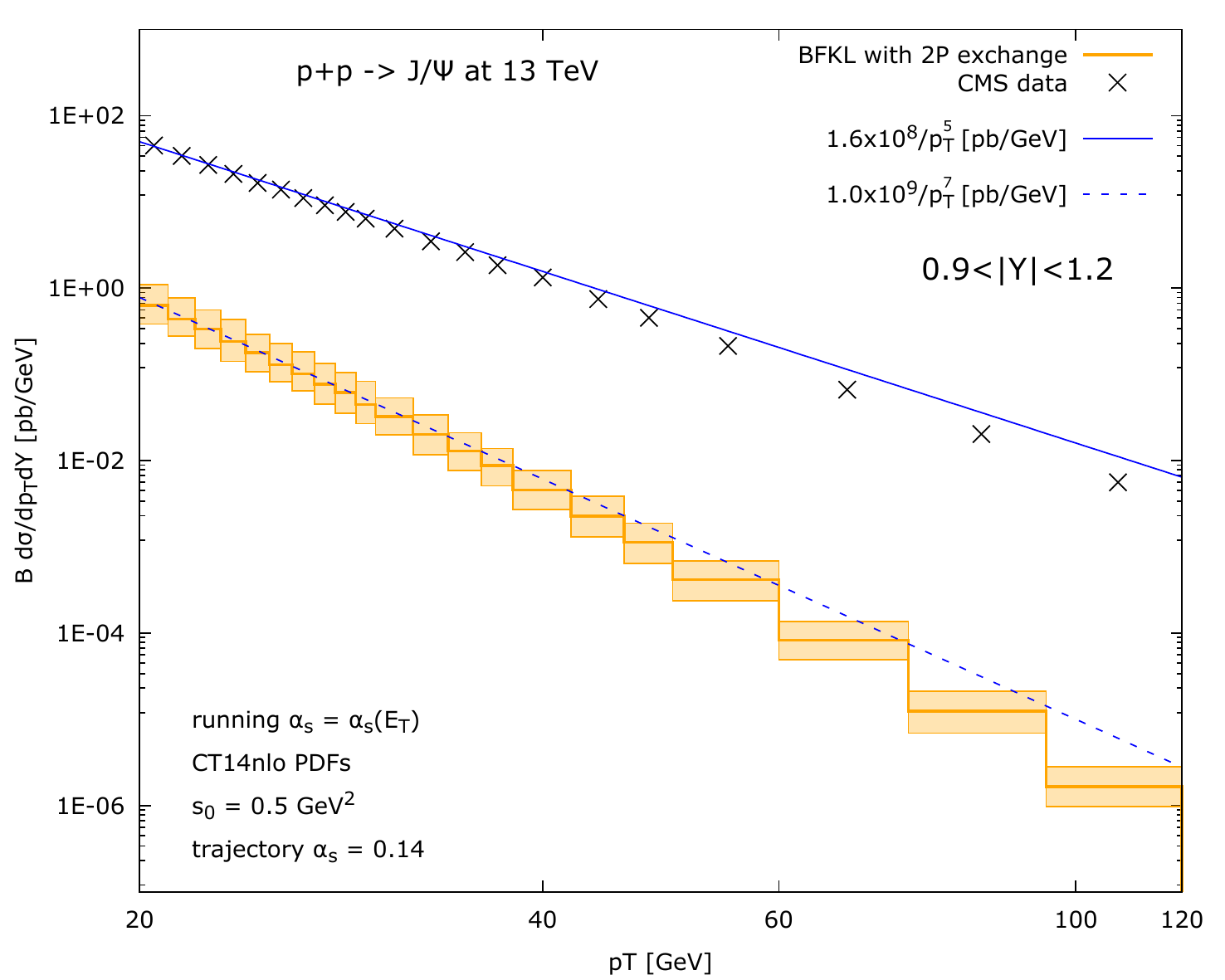}}
\caption{The double differential cross section $d\sigma/dp_TdY$  for the hadroproduction of the $J/\psi$, times the branching ratio for leptons $B$. Here $Y$ is the rapidity of the $J/\psi$. The solid histograms represent the double-Pomeron contribution to the cross section calculated in this work.  The error estimate (shaded boxes) comes from the scale variation. We also show the power-law behaviour $C_1/p_T^5$ and $C_2/p_T^7$. }
\label{fig:result_hadroprod1}
\end{figure}

Let us now turn to the double-Pomeron contribution to hadroproduction of $J/\psi$ and $\Upsilon$.
The point we want to focus on is the $p_T$-dependence of the vector meson in hadroproduction at large $p_T$. This is because the various meson production mechanisms are characterized by distinct and robust resulting $p_T$ dependencies. This allows for the most stringent verification of the production mechanism and so far provides the strongest evidence for the dominance of the COM.
Therefore we choose for the comparisons of the theoretical predictions with the experimental results, the dataset with the largest available $p_T$ range. These are measurements of prompt $J/\psi$ and $\Upsilon$ production by the CMS collaboration in $pp$ collisions at $\sqrt{S}=13$~TeV  \citep{Sirunyan:2017qdw}. The kinematic setup for our calculation is adjusted to match this dataset.  
Accordingly, we set the c.m.s.\ proton--proton collision energy to 13~TeV and collect the Monte Carlo events in 2D bins of $p_T$ and rapidity $Y$ of the vector meson reconstructed as
\begin{equation}
Y = \log \left( \frac{x_1 \sqrt{S}}{E_T} \right)\, .
\end{equation}
 Other parameters of the calculation are unchanged with respect to the diffractive photoproduction case (obviously, except the mass of the vector meson for the hadroproduction of $\Upsilon (1S)$ state).
We make the slices on the 2D histogram in $|Y|$ following the CMS setup. Comparison of the two-Pomeron contribution to $J/\psi$ hadroproduction and the respective data for four rapidity slices is shown in Fig.~\ref{fig:result_hadroprod1}.

\begin{figure}[t]
\centerline{\includegraphics[width=10cm,angle=0]{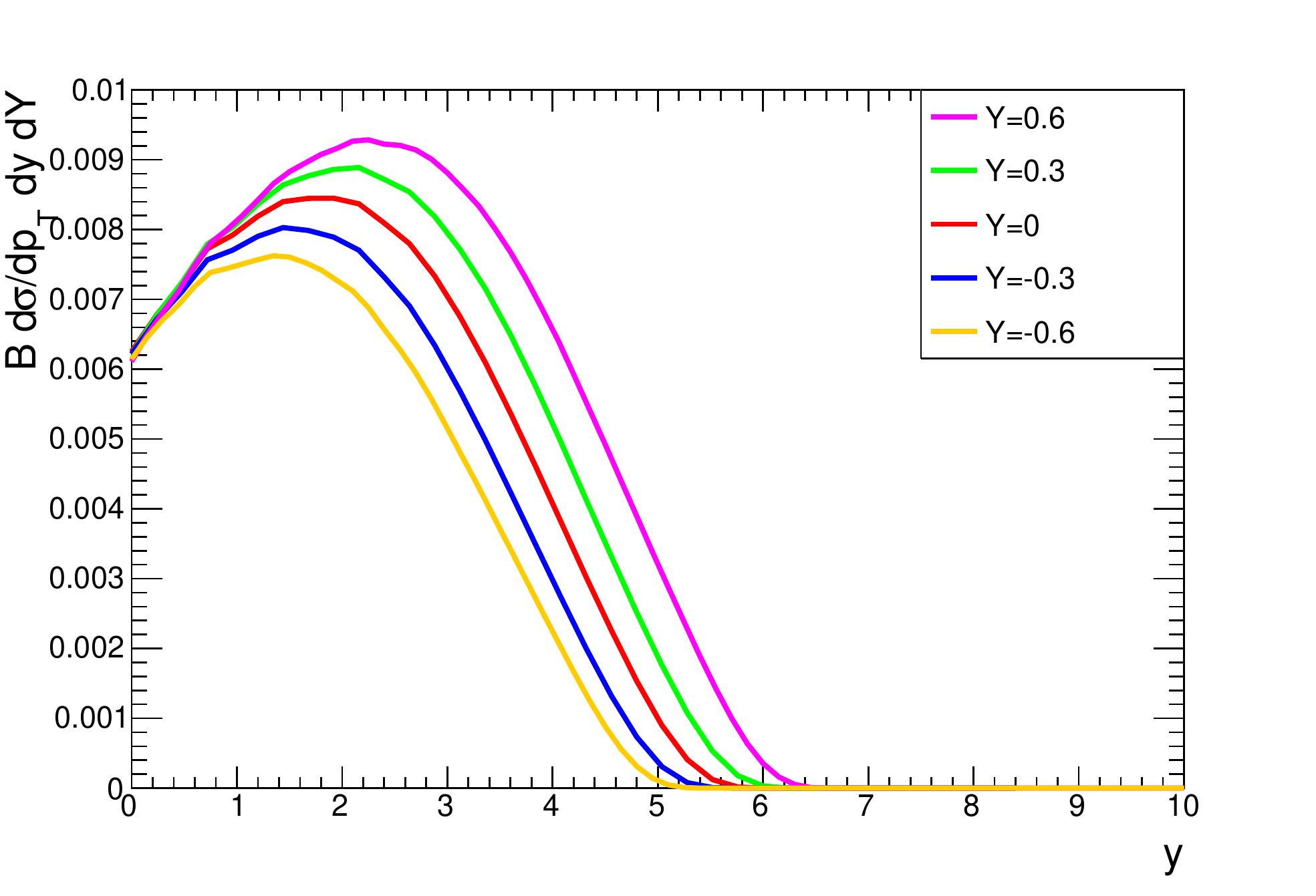}}
\caption{
Distributions of the BFKL evolution length $y$ in $J/\psi$ production in proton--proton collisions by correlated double Pomeron exchange for $\sqrt{S}=13$~TeV, $p_T =30$~GeV and $Y=-0.6,-0.3,0, 0.3, 0.6$. Only   the first term (asymmetric contribution) in Eq.\eqref{eq:HadroXsec1} was included to better illustrate the dependence on the length of the rapidity evolution in the BFKL Pomeron.  }
\label{fig:result_hadroprod_y0}
\end{figure}

The first striking observation is that the double-Pomeron contribution is very small in the kinematic region probed. 
On one hand, this is something one could somewhat naively expect from the Born level formula Eq.~(\ref{eq:HadronBorn}) due to the high power of strong coupling constant in the cross section, i.e.\ $\alpha_s^5$. On the other hand, in principle, the Born cross section could gain strong enhancement due to the double BFKL evolution. 
It turns out, however, that the evolution length for the kinematics tested experimentally is not sufficiently long to compensate for the $\alpha_s^5$ suppression factor.
To demonstrate this we calculate the distributions of the rapidity distance~$y$,
\begin{equation}
y=\log \left( \frac{\hat{s}}{E_T^2} \right) \, ,
\end{equation}
which actually sets the evolution length of the BFKL Green's functions.
The results for the $y$-distribution in $J/\psi$ proton--proton collisions at $\sqrt{S} = 13$~TeV for typical meson $p_T$ = 30 GeV and rapidities $Y = -0.6,-0.3,0, 0.3,0.6$ are shown in Fig.~\ref{fig:result_hadroprod_y0}.  We note that, in this plot we only took one contribution to the cross section, i.e.\ the first term in Eq.\eqref{eq:HadroXsec1}.  This was done in order to better illustrate the shift of the peak of the distributions with the changing rapidity of the vector meson, due to the change of the BFKL evolution. Of course in the full calculation the symmetry with respect to exchange of $x_1$ and $x_2$ is taken into account.
The distributions peak around $y \sim 2$, and decrease towards the kinematical cutoff at $y \sim 6$ (depending on $Y$). The mean evolution length in $y$ is about 2 units, which is rather short and implies only moderate BFKL enhancement. 
A naive estimate using asymptotic enhancement factors of two the BFKL Pomeron exchange gives $\exp(2\Delta_{\mathbb{P}} y) \sim 4.5$ for $y=2$ and the LL BFKL intercept $\Delta_{\mathbb{P}} \simeq 0.39$ corresponding to the assumed effective value $\bar\alpha_s = 0.14$ in the BFKL kernel. It is well known however, that preasymptotic effects lead to a decrease of the enhancement w.r.t.\ its asymptotic form. 

\begin{figure}[t]
\centerline{\includegraphics[width=10cm,angle=0]{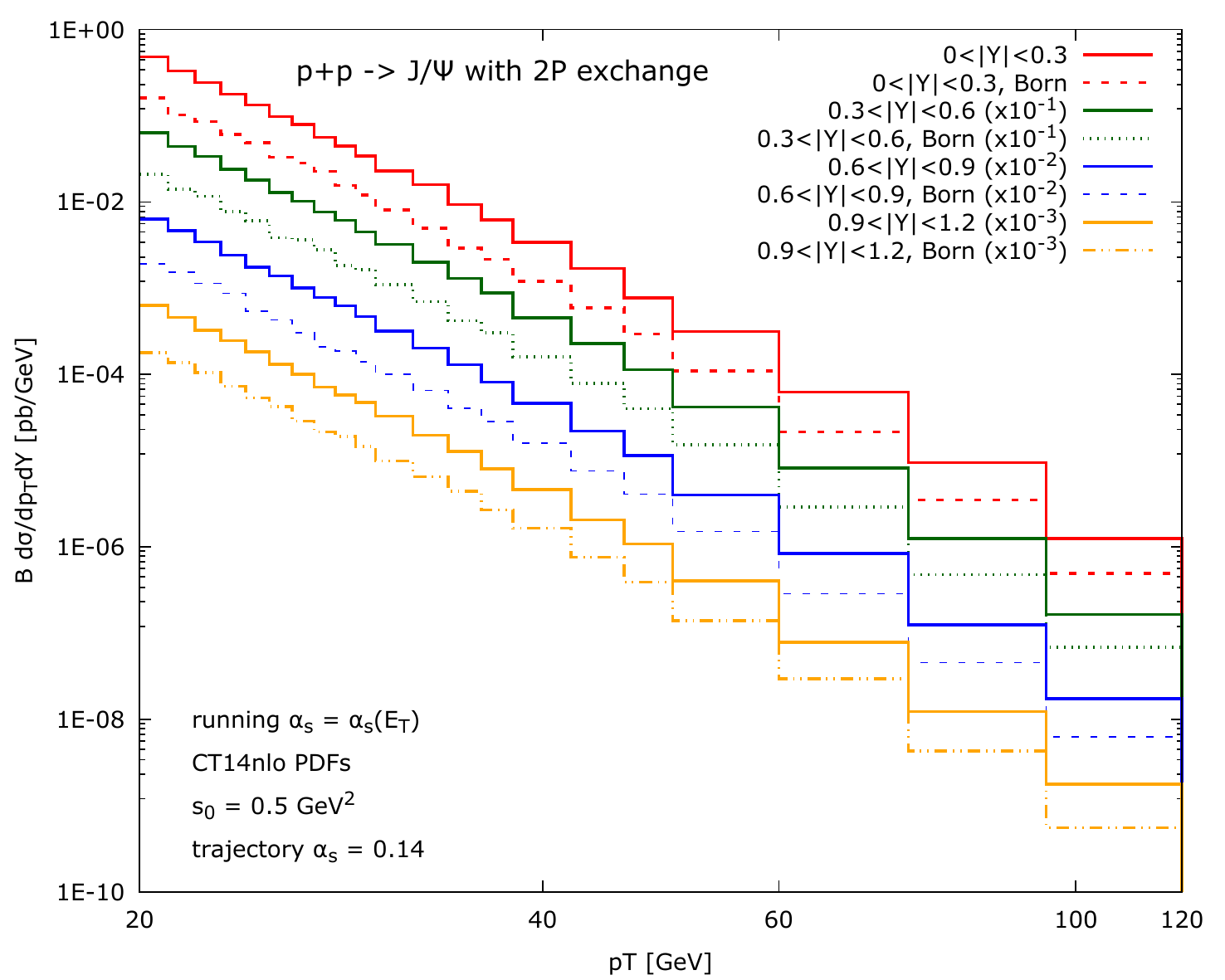}}
\caption{The same theoretical distributions as in Fig.~\ref{fig:result_hadroprod1} but gathered on single
plot and compared with the Born-level cross section.
}
\label{fig:result_hadroprod_born}
\end{figure}

The actual enhancement due to the evolution of the double BFKL ladder is shown in Fig.~\ref{fig:result_hadroprod_born}. We see that indeed the gain is roughly a factor of two. The moderate length $y$ of the BFKL evolution may be understood as result of a competition between $x$-dependencies of the PDFs that weight the partonic correlated 3GF cross section of the vector meson production off a quark or gluon, and the double BFKL evolution. Indeed, the PDFs grow towards small $x$ as $x^{-1-\lambda}$ with $\lambda \sim 0.3$, so they strongly push towards small $x_i$ values of the partons. The PDF factor $\sim (x_1 x_2)^{-1-\lambda}$  squeezes the phase space for the partonic process and leads to only moderate average $y$.

\begin{figure}[t]
\centerline{\includegraphics[width=10cm,angle=0]{{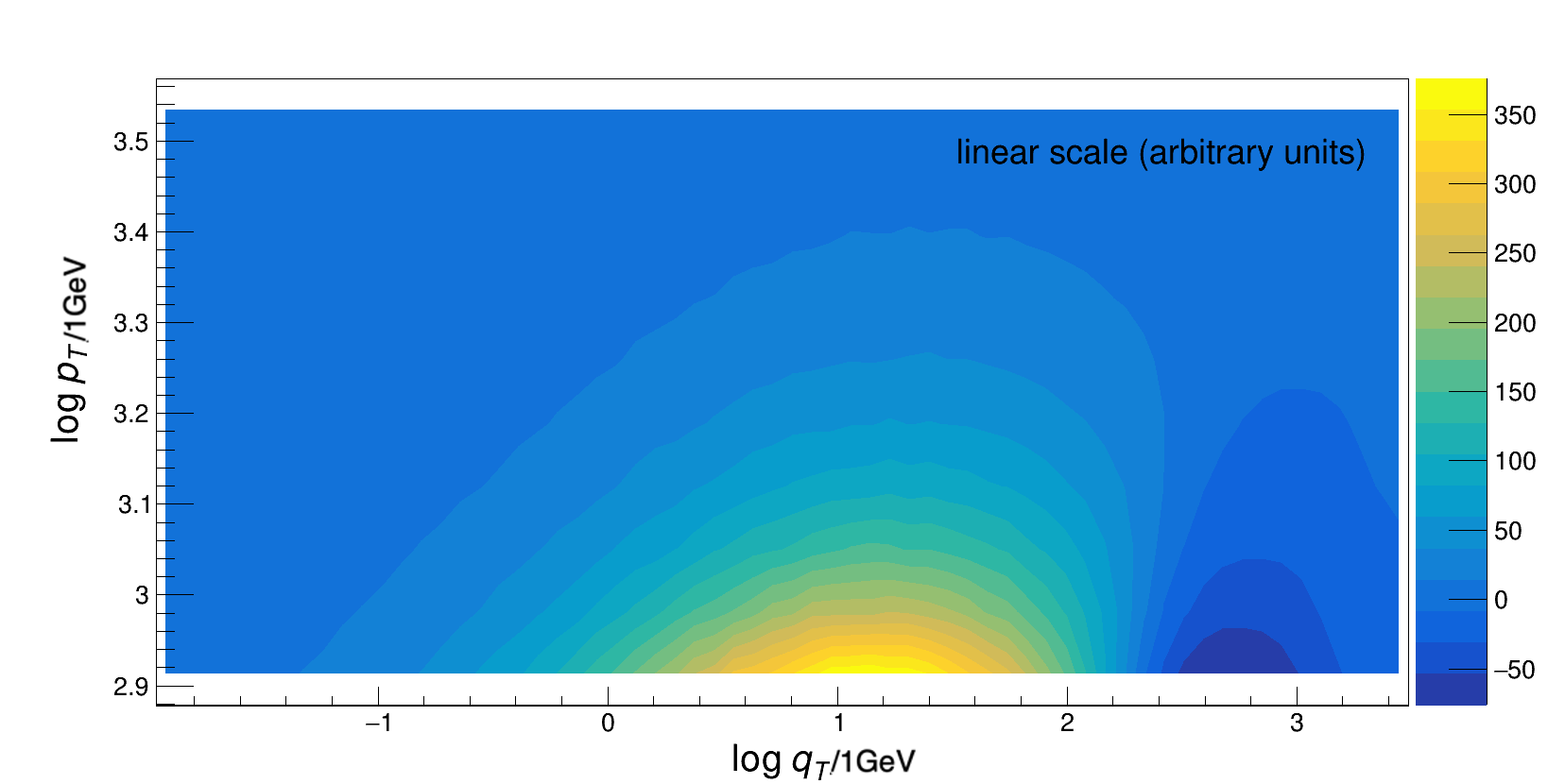}}}
\caption{The hadroproduction cross section (in arbitrary units) as a function of the transverse momentum of the vector meson 
$\boldsymbol{p}$ and the
transverse momentum flowing in the double-Pomeron loop $\boldsymbol{q}$, for the kinematics
probed by the CMS measurement. The deep blue region gives a negative contribution to the integral over $\boldsymbol{q}$, but the cross section differential in physically accessible variable $|\boldsymbol{p}|$ remains positive. 
}
\label{fig:result_hadroprod_qT2}
\end{figure}

We also study the hadronic cross section as a function of the transverse momentum $\boldsymbol{q}$ flowing in the double-Pomeron loop. In Fig.~\ref{fig:result_hadroprod_qT2} we plot the contribution to the cross section, unintegrated in both $p_T$ and $q_T$, similar to what has been done in Fig.~\ref{fig:PartXsec2D} but at the hadronic level and within the kinematic window studied by the CMS experiment. We observe a non-negligible negative contribution to the integral coming from regions of non-zero momentum transfer flowing in the Pomeron loop. This is an exclusive property of the correlated two Pomerons and constitutes an additional source of the weaker cross section enhancement than naively expected. 
Note, that for any $|\boldsymbol{p}|$ slice the cross section will turn into a positive number upon integration over $\boldsymbol{q}$.

The second observation regarding the comparison of the calculation with the experimental results in Fig.\ \ref{fig:result_hadroprod1} is the power-like behavior. For the data
we observe roughly $1/p_T^5$ fall-off, while the double-Pomeron contribution dies out faster, roughly like $1/p_T^{7+\epsilon}$ with $\epsilon <1$. 
The obtained $p_T$-dependence at the hadron level is significantly steeper than the parton level cross section Eq.~(\ref{eq:HadronBorn}) with a fixed coupling constant. This happens mostly because the strong dependence of the hadronic cross section to kinematic lower cutoffs $x_i ^{\mathrm{min}}$ on the parton $x$ values. One obtains
 $x_i ^{\mathrm{min}} \sim E_T / \sqrt{S}$ and these constraints lead to steeper dependence of the hadronic cross sections with $E_T$. In addition $\alpha^5 _s(E_T)$ in the impact factors leads to an additional suppression with increasing $E_T$.

\begin{figure}[t]
\centerline{\includegraphics[width=8cm,angle=0]{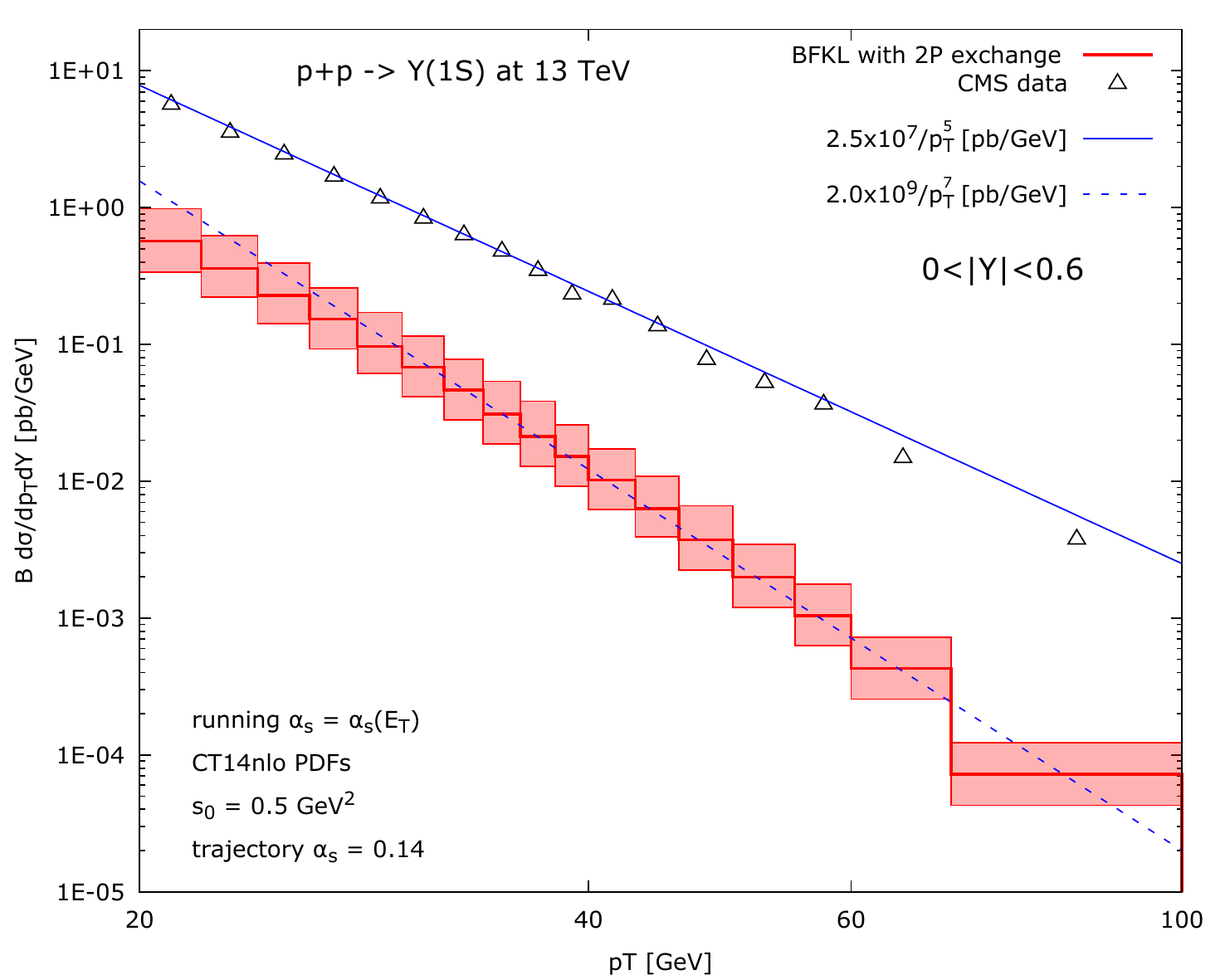}\quad
\includegraphics[width=8cm,angle=0]{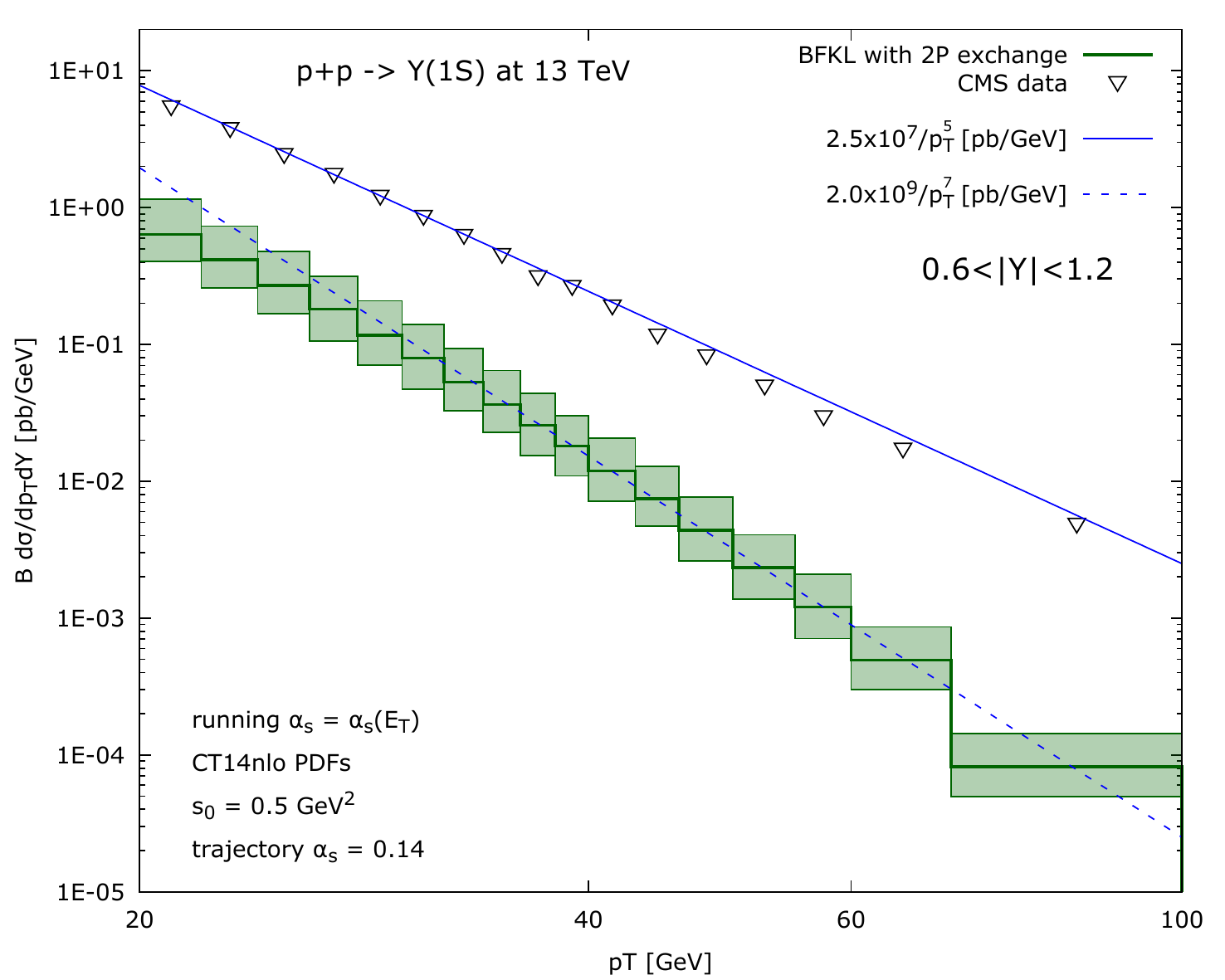}}
\caption{Same as in Fig.~\ref{fig:result_hadroprod1} but for hadroproduction of $\Upsilon (1S)$. }
\label{fig:result_hadroprod2}
\end{figure}

For completeness, we also perform the calculation for the $\Upsilon$(1S) case. The result is presented in Fig.~\ref{fig:result_hadroprod2}. For this case, all the previous observations apply, except the power-like scaling of the
correlated double-Pomeron contribution, which is now less accurate at lower values of $p_T$. This is expected as in this region the effects of the second scale of the problem --- the meson mass $M_V$ become relevant for $\Upsilon$, that is much heavier than $J/\psi$.

\newpage

\section{Discussion}
\label{sec:Discussion}

The analyses performed in this paper are based on the non-forward BFKL equation in the leading logarithmic approximation. The NLL corrections to the BFKL kernel are known to affect strongly the rapidity dependence of the amplitudes. We represent these effects by using an effective small value of the fixed coupling constant in the LL BFKL kernel. This approach is then confronted with the diffractive photoproduction data. The present paper is focused on estimating the correlated two Pomeron contribution to the inclusive hadroproduction of heavy vector mesons. 
Therefore we use the diffractive photoproduction data to constrain the framework, and then apply it to the hadroproduction. We show that the $J/\psi$ diffractive photoproduction data may be well described within the LL BFKL approach, so we do not find it necessary to consider the diffractive $J/\psi$ photoproduction at large $p_T$ within the
full next-to-leading logarithmic BFKL approximation. The tuned LL BFKL framework applied to inclusive hadroproduction gives a small contribution to the differential cross sections for $J/\psi$ and $\Upsilon$ hadroproduction at large $p_T$ and does not explain the meson hadroproduction data. These main conclusions are not sensitive to details of the description. With this outcome it is not necessary to treat subleading effects, like the real parts of the BFKL exchanges or corrections due to off-diagonal effects --- e.g.\ the Shuvaev factor \cite{Shuvaev:1999ce}. In the present approach such correction factors may be
 efficiently absorbed into the effective value of $\alpha_s$ driving the BFKL evolution.

The strongest source of the theoretical uncertainty in the obtained results is the scale $\mu$ dependence of $\alpha_s(\mu)$ in the impact factors. The diffractive photoproduction cross section is proportional to $\alpha_s ^4(\mu)$, and the analyzed triple gluon fusion contribution to the hadroproduction is proportional to $\alpha_s ^5(\mu)$. Using the standard procedure of varying $\mu$ between the half and double of the natural process scale $\mu_0 = E_T$ one finds rather large factor of about~2 (or more) of theoretical uncertainty due to the scale variation, see Fig.\ \ref{fig:result_diffraction1} for diffractive photoproduction, and similarly for the hadroproduction. These uncertainty factors are, however, correlated at different values of $p_T$ and they affect the overall normalization of the differential cross sections more than the shapes. Hence the agreement of the theoretical predictions and the data for the $p_T$-shape of the diffractive photoproduction is rather robust. On the contrary --- for the inclusive hadroproduction the correlated two Pomeron exchange leads to a much too steep $p_T$~dependence.

\begin{figure}[t]
\centerline{\includegraphics[width=10cm,angle=0]{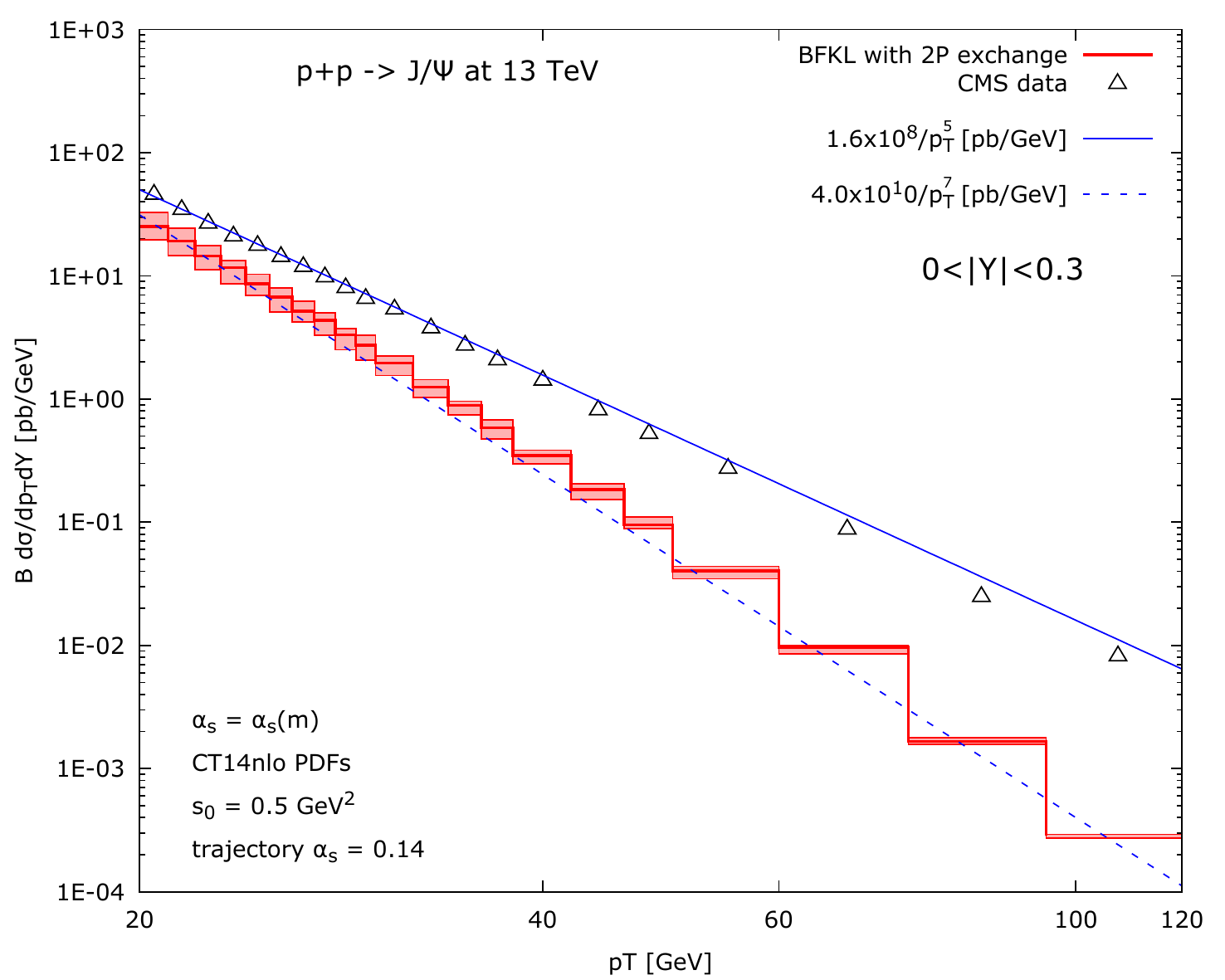}}
\caption{The $J/\psi$ cross section obtained with the fixed coupling constant $\alpha_s(m_c)$ in the impact factors.
}
\label{fig:result_hadroprod_y}
\end{figure}

In a recent analysis \cite{Levin:2018qxa} a good description of the direct $J/\psi$ production in $pp$ collisions at the LHC was obtained with the uncorrelated 3GF mechanism, without any other contributions included. Those results differ significantly from ones found in Ref.\ \cite{Motyka:2015kta}. The main reasons of the difference is using $\alpha_s(m_c)$ in the impact factors in Ref.\ \cite{Levin:2018qxa}, while in  Ref.\ \cite{Motyka:2015kta} the running coupling constant $\alpha_s(E_T)$ was used. This leads to a modification of the obtained transverse momentum distribution, because of a large span of the transverse momentum probed and the $\alpha_s^3$ dependence of the cross section. The choice of the scale  $\mu = E_T$ is also applied in the current paper, as it is related to the highest virtuality that occurs in the vector meson production vertex. In order to test the sensitivity of the results to the different choice of scale we  also compute  the cross section for the correlated 3GF mechanism of $J/\psi$ hadroproduction in which the fixed $\alpha_s(m_c)$ is used. The results are shown in Fig.\ \ref{fig:result_hadroprod_y} for the central rapidity bin $|Y|<0.3$ . The effect of the change from   $\alpha_s(\mu)$ to $\alpha_s(m_c)$ is dramatic, as expected from the $\alpha_s^5$ dependence of the cross section. The obtained cross section is closer to the data at lower $p_T \sim 20$~GeV. The $p_T$ dependence is harder than it is for the running coupling, but still somewhat softer than the data. 

Let us finally make connection of the computed amplitudes to the  triple BFKL Pomeron vertex  derived by Bartels \cite{Bartels:1992ym,Bartels:1993ih}. In our approach two BFKL Pomerons are coupled to a single parton at the target side. The parton emerges from the target and undergoes the  QCD evolution, which in the small~$x$ domain may be also described by the BFKL equation. For the gluon parton, this topology corresponds to some of the diagrams that enter the triple Pomeron vertex. In the computation we apply the Mueller--Tang approximation, that isolates the dominant contribution to the  amplitude. A more complete treatment of this configuration should be possible using the results obtained for all cut triple Pomeron vertices \cite{Bartels:2008ru}.

\section{Summary and conclusions}
\label{sec:Summary}

In this paper we have performed a detailed analysis of  the contribution to the vector meson hadroproduction due to the correlated three gluon fusion mechanism (3GF).  In this process, a vector meson, $J/\psi$ or $\Upsilon$ in our analysis, is  directly formed as a color  singlet state from the interaction of three gluons. To be precise, in the process considered, one gluon originates from one hadron and two other gluons from another hadron. In the particular case of the correlated  3GF, we   considered the contribution when two gluons  came from the perturbative splitting of a parent quark or a gluon. 
The produced vector meson and the projectile quark and gluon were separated by a  rapidity distance $y$ which we modeled as an exchange of two cut non-forward BFKL Pomerons in the leading logarithmic approximation at high energy.  Effectively this process contains a Pomeron loop contribution which is spanned between one collinear parton from one of the  hadrons and the vector meson impact factor.  For consistency and in order to fix the parameters as well as the normalization we have simultaneously calculated the diffractive photoproduction of vector mesons in deep inelastic scattering through the mechanism of exchange of the same BFKL Pomerons. The topology of this process, modulo color factors, is the same as the correlated 3GF vector meson hadroproduction. The important difference lies in the different cuts through the BFKL Pomerons: in the case of the diffractive DIS, one considers the diffractive cut, which leads to the rapidity gap between the vector meson and the target remnant, and in the case of the hadroproduction both BFKL Pomerons are cut. The appropriate color factors for the case of vector meson hadroproduction were obtained by the projection of the BKP states
onto the color combination corresponding to two cut Pomerons  and retaining terms leading in $N_c$. 
The correlated Pomeron loop was  numerically first  analyzed at the level of the partonic cross section as a function of the transverse momentum of the produced vector meson and the rapidity. We found that, the cross section exhibits a dip in the transverse momentum, at the level of the Born cross section (i.e.\ without the BFKL evolution) which occurs at about the value corresponding to the vector meson mass.
When the BFKL evolution is included, the dip is washed out and almost completely disappears for high values of rapidity. We checked that the rapidity dependence of the partonic cross section is consistent with the one expected from the two BFKL Pomerons, i.e.\ the rapidity dependence is exponential with the intercept that corresponds to twice the value of  the intercept of the BFKL Pomeron. 
The momentum transfer dependence inside the Pomeron loop was also analyzed, and it turns out that the distribution in this variable has a sharp cutoff at the values which correspond to approximately half of the transverse momentum of the vector meson, i.e.\ $q_T\sim p_T/2$, after which the distribution becomes slightly negative.
This cutoff stems from the coupling of two Pomerons to the vector meson impact factor.

We have then evaluated the cross section for the diffractive photoproduction of $J/\psi$ in DIS, for large values of momentum transfer.  The non-forward BFKL evolution was solved numerically for fixed values of the strong coupling constant. Since the NLL corrections to the BFKL evolution are known to be large we have used the value of the strong coupling as a fitting parameter. The other freedom was in the choice of the infrared cutoff (mimicking the effective gluon mass due to the color confinement)
inside the BFKL evolution and the choice of the scale in the running coupling in the  impact factors. We also computed the cross section weighted by the equivalent photon flux. We obtained very good agreement with the experimental data from H1 collaboration as a function of the momentum transfer. 
Using the same parameters then, we have obtained essentially parameter free prediction for the 3GF correlated contribution to the vector meson hadroproduction. We made the comparison with the CMS data for $J/\psi$ and $\Upsilon$ production at $\sqrt{S}=13\, \rm TeV$ in bins of rapidity of the produced vector meson and as a function of the transverse momentum.    We have found that this contribution is rather small, of the order of few percent at smallest values of the transverse momenta. The magnitude of the cross section is only slightly enhanced by the BFKL growth as compared with the Born cross section with just four gluon exchange, despite the fact that two Pomerons are exchanged. We further analyzed this, by evaluating the typical values of the rapidities probed in the CMS kinematics, and found that indeed the cross section is dominated by rather low rapidity intervals, and that the BFKL Pomeron exchange leads to about factor two enhancement over the Born cross section.  The correlated 3GF contribution to the cross section falls off like $1/p_T^7$, which is faster than the $1/p_T^5$ seen in the data, and is consistent with the evaluation of the Born cross section.

Several improvements could be made to the calculation presented in this paper. Non-forward BFKL evolution is known to NLL order, and this could be used to estimate better the value of the cross section. Combined with the NLL impact factors, this would be the first complete NLL evaluation of the correlated Pomeron loop contribution to this process.

\section*{Acknowledgments}
We thank Jochen Bartels for valuable comments and stimulating discussions.
AMS was supported by the  Department of Energy Grant No. DE-SC-0002145, as well as the National Science centre, Poland, Grant No.\ 2015/17/B/ST2/01838. Support of the NCN grant No.\ 2017/27/B/ST2/02755 is gratefully acknowledged.
\appendix

\newpage
\bibliographystyle{JHEP}
\bibliography{references}

\end{document}